\documentclass[acmsmall]{acmart}

\AtBeginDocument{%
  }

\usepackage{xcolor}
\definecolor{darkgreen}{rgb}{0.0, 0.4, 0.3}

\usepackage{mathtools}
\usepackage{multirow}
\usepackage{natbib}
\usepackage{amsmath}
\usepackage{caption}
\usepackage{comment}
\usepackage{longtable}
\usepackage{multicol}
\usepackage[export]{adjustbox}
\usepackage{xcolor}
\usepackage{longtable,rotating}
\usepackage[normalem]{ulem}

\setcopyright{cc}
\setcctype{by}
\acmJournal{PACMHCI}
\acmYear{2025} \acmVolume{9} \acmNumber{7} \acmArticle{CSCW476} \acmMonth{11} \acmPrice{}\acmDOI{10.1145/3757657}

\newcommand{\addNew}[1]{#1}  
\newcommand{\delNew}[1]{}   

\newcommand{\add}[1]{#1}  
\newcommand{\del}[1]{}   

\begin{document}

\title{Signals in the Noise: Decoding Unexpected Engagement Patterns on Twitter}


\author{Yulin Yu}
\email{yulinyu@umich.edu}
\orcid{0000-0000-0000-0000}
\affiliation{%
  \institution{University of Michigan}
  \city{Ann Arbor}
  \state{Michigan}
  \country{USA}
  \postcode{48109-1285}
}

\author{Houming Chen}
\email{houmingc@umich.edu}
\orcid{0000-0000-0000-0000}
\affiliation{%
  \institution{University of Michigan}
  \city{Ann Arbor}
  \state{Michigan}
  \country{USA}
  \postcode{48109-1285}
}

\author{Daniel Romero}
\email{drom@umich.edu}
\orcid{0000-0000-0000-0000}
\affiliation{%
  \institution{University of Michigan}
  \city{Ann Arbor}
  \state{Michigan}
  \country{USA}
  \postcode{48109-1285}
}

\author{Paramveer S. Dhillon}
\email{dhillonp@umich.edu}
\orcid{0000-0000-0000-0000}
\affiliation{%
  \institution{University of Michigan}
  \city{Ann Arbor}
  \state{Michigan}
  \country{USA}
  \postcode{48109-1285}
}

\renewcommand{\shortauthors}{Yulin Yu, Houming Chen, Daniel Romero, and Paramveer S. Dhillon}


\begin{abstract}
Social media platforms offer users multiple ways to engage with content—likes, retweets, and comments—creating a complex signaling system within the attention economy. While previous research has examined factors driving overall engagement, less is known about why certain tweets receive unexpectedly high levels of one type of engagement relative to others. Drawing on Signaling Theory and Attention Economy Theory, we investigate these unexpected engagement patterns on Twitter\footnote{The social media platform has been renamed to `X,' however our study data is from the time-period prior to the name change.}, developing an ``unexpectedness quotient'' to quantify deviations from predicted engagement levels. Our analysis of over 600,000 tweets reveals distinct patterns in how content characteristics influence unexpected engagement. News, politics, and business tweets receive more retweets and comments than expected, suggesting users prioritize sharing and discussing informational content. In contrast, games and sports-related topics garner unexpected likes and comments, indicating higher emotional investment in these domains. The relationship between content attributes and engagement types follows clear patterns: subjective tweets attract more likes while objective tweets receive more retweets, and longer, complex tweets with URLs unexpectedly receive more retweets. These findings demonstrate how users employ different engagement types as signals of varying strength based on content characteristics, and how certain content types more effectively compete for attention in the social media ecosystem. Our results offer valuable insights for content creators optimizing engagement strategies, platform designers facilitating meaningful interactions, and researchers studying online social behavior.
\end{abstract}



\received{October 2024}
\received[revised]{April 2025}
\received[accepted]{August 2025}

\maketitle

\section{Introduction}

\add{Social media platforms have become central arenas for information exchange, social interaction, and content dissemination \citep{boyd2010tweet,zhang2018characterizing,burke2010social}. The unique affordances of these platforms, such as likes, retweets, and comments, have created a complex ecosystem of user engagement \citep{meier2014more,scissors2016,gilbert2013widespread}. While previous studies have analyzed overall engagement, we examine {\it unexpected engagement,} defined as situations in which content receives disproportionately high levels of one specific interaction type (likes, retweets, or comments) relative to the others. For example, a tweet providing in-depth political analysis might receive an unexpectedly high number of retweets compared to likes or comments, as users prioritize information dissemination over expressing agreement \citep{gonzalez2010structure,zhang2015modeling,semaan2014social}. Conversely, controversial opinions often generate an unusually high volume of comments relative to likes or retweets, as users engage in debate without endorsing the message \citep{zhang2018characterizing,grevet2014managing,semaan2015navigating}. Similar patterns emerge with emotional content, where personal narratives tend to accumulate a disproportionate number of likes compared to other forms of engagement, as users express support without feeling compelled to share or discuss further \citep{levordashka2016s,burke2016relationship,luo2020self}.}

\add{A substantial body of research has examined social media engagement mechanisms, revealing common motivations and patterns behind user behaviors across platforms~\citep{levordashka2016s,boyd2010tweet,meier2014more}. While several studies have analyzed engagement metrics collectively~\citep{tenenboim2015prompts,aldous2019view,pancer2016popularity}, they have primarily focused on predicting overall popularity or engagement volumes rather than examining the relationships between different engagement types. Previous research has identified key content features driving engagement: topic emerges as a primary factor, with domains like music, sports, and news generating distinctive engagement profiles~\citep{meier2014more,romero2011differences,gonzalez2010structure}; emotional valence significantly affects interaction patterns~\citep{meier2014more,berger2012makes,stieglitz2013emotions,tan2014effect}; and textual complexity shapes user responses through factors like readability, URL inclusion, and content length~\citep{boyd2010tweet,meier2014more,suh2010want,tan2014effect}. Our work extends these findings by examining how these content characteristics drive unexpected deviations in engagement composition rather than just volume. Unlike previous studies that treat engagement as a relatively uniform construct with varying intensities, we demonstrate how different content features systematically elicit specific types of unexpected engagement, revealing distinct signaling behaviors and attention allocation patterns that challenge existing engagement models.}

\add{While these studies have provided valuable insights into how individual features drive overall engagement, they have not explored how these factors might contribute to asymmetric interaction patterns\textemdash situations where content receives unexpectedly high levels of one type of engagement relative to others. This gap is particularly significant as social media platforms increasingly serve as crucial channels for information dissemination and public discourse. To bridge this gap and provide deeper theoretical insights, we draw on two complementary frameworks: {\it Signaling Theory} and {\it Attention Economy Theory.}}

\add{Signaling Theory and Attention Economy Theory together provide a comprehensive framework for analyzing unexpected engagement patterns. Originally developed to explain animal communication through observable actions \citep{donath2007signals,connelly2011signaling}, Signaling Theory illuminates how users leverage different engagement mechanisms as signals with varying strength and meaning. In social media contexts, likes function as low-cost signals of approval, retweets represent stronger endorsement as users share content with their networks \citep{sharma2012inferring,bernstein2013quantifying}, and comments—requiring the most cognitive effort—demonstrate a willingness to invest time and thought, constituting the strongest form of engagement \citep{gilbert2013widespread,burke2010social}. Complementing this, Attention Economy Theory posits that in information-rich environments, the ability to attract and maintain attention becomes a critical success factor \citep{davenport2001attention,webster2014marketplace}. On platforms like Twitter, where users navigate constant information overload, content must compete effectively for limited attention resources to avoid being lost in the endless scroll \citep{mahmud2013will,wang2014whispers}.}

\add{Our study makes significant theoretical contributions by synthesizing these frameworks to develop a novel methodological approach\textemdash the ``unexpectedness quotient''\textemdash that quantifies deviations from expected engagement patterns. This interdisciplinary approach bridges insights from communication studies, psychology, and information science \citep{bernstein2013quantifying,burke2010social}, enabling us to examine not just overall engagement volume but the specific composition of engagement types. While previous research has explored general patterns, our study focuses on understanding and predicting unexpected deviations in engagement ratios, providing new insights into the nuanced ways users interact with social media content \citep{scissors2016}. This approach allows us to identify content characteristics that systematically drive specific types of unexpected engagement, revealing patterns of user behavior that challenge conventional engagement models. Building on this novel methodological approach, we specifically seek to answer the following research questions:}

\begin{itemize}
\item {\bf RQ1:} How can we quantify and measure unexpected engagement patterns across different interaction types (likes, retweets, and comments)?
\item {\bf RQ2:} Which content attributes account for deviations from expected interaction patterns, and to what extent do features such as textual complexity, sentiment, topic, subjectivity, and the presence of links predict the unexpectedness of a tweet's engagement patterns?
\end{itemize}

\add{To investigate these research questions empirically, we analyze a comprehensive dataset of over 600,000 tweets using the ``unexpectedness quotient'' that we develop. This approach allows us to identify tweets that receive disproportionate levels of specific engagement types\textemdash for example, content that generates significantly more comments than its like or retweet counts would predict. We examine how various content characteristics influence these unexpected engagement patterns, including textual complexity (readability, length, URL presence), valence (sentiment, subjectivity), and content topic, while controlling for author attributes. This large-scale analysis reveals systematic relationships between specific content features and particular types of unexpected engagement that challenge conventional understanding of social media interaction patterns \citep{gilbert2013widespread}.}

\add{Our analysis reveals several key findings that illuminate the complex dynamics of unexpected engagement patterns on social media. The topic of the content strongly influences the type of engagement that predominates: news, politics, and business-related tweets consistently receive more retweets and comments than expected, while games and sports-related topics generate unexpected likes and comments. Art-related content primarily attracts disproportionate likes, while celebrity and movie topics tend to garner both unexpected likes and retweets. Examining content characteristics beyond topic, we find that subjective content systematically attracts more likes than predicted, while objective content receives unexpected retweets. Tweets containing URLs and those with greater textual complexity receive more retweets than expected, suggesting that information-rich content drives sharing behavior. The presence of concrete language correlates with higher-than-expected likes, while longer tweets generate more unexpected comments, indicating that content length may stimulate discussion. These patterns challenge existing engagement models that treat user interactions as uniform behaviors that differ merely in intensity \citep{tenenboim2015prompts,aldous2019view,pancer2016popularity}, revealing instead systematic relationships between specific content characteristics and particular types of engagement.}

\add{These findings substantially extend both Signaling Theory and Attention Economy Theory in digital contexts. Traditional applications of Signaling Theory treat platform affordances (likes, retweets, comments) as signals with fixed strength hierarchies, but our results demonstrate that signal utilization varies systematically based on content characteristics across different domains \citep{connelly2011signaling, donath2007signals, sharma2012inferring}. Users strategically select engagement mechanisms based on their specific communicative function within particular content domains\textemdash employing high-cost signals for informational content while balancing low-cost signals with community participation in emotionally engaging contexts. Similarly, our results advance Attention Economy Theory beyond treating attention as a uniform resource \citep{davenport2001attention, goldhaber1997attention, webster2014marketplace}, revealing qualitatively different forms of attention investment. Users allocate distribution attention (retweets) toward objective, informational content while investing evaluative attention (likes) in subjective content, directly contradicting conventional models that aggregate interaction types into composite ``engagement scores'' \citep{sekimoto2020metrics,pancer2016popularity}. The prevalence of unexpected comments in certain contexts suggests that specific content qualities trigger deliberative attention investment without necessarily generating viral spread \citep{burke2016relationship,mahmud2013will,scissors2016}. These patterns reveal social media attention as a complex ecosystem where content characteristics compete for qualitatively distinct forms of attention investment, significantly advancing beyond current engagement models that fail to differentiate between interaction types.}

\del{Our work offers valuable insights for content creators seeking to optimize specific types of engagement, platform designers aiming to facilitate more meaningful interactions, and researchers studying online social behavior. As social media continues to shape public discourse, personal relationships, and communication strategies, understanding the factors that drive different types of engagement becomes increasingly crucial \citep{burke2010social}. Our findings not only advance theoretical understanding of user behavior in digital spaces but also offer practical insights for developing more effective strategies in the complex landscape of social media engagement \citep{bernstein2013quantifying}. This research has implications for improving content recommendation systems, enhancing user experience design, and developing more sophisticated social media analytics tools that consider the multifaceted nature of user engagement.}

\add{These theoretical extensions to Signaling Theory and Attention Economy Theory translate directly into actionable insights for stakeholders across the social media ecosystem. For platform designers, our findings on unexpected engagement patterns could inform more sophisticated recommendation algorithms that go beyond maximizing overall engagement volume to promoting specific types of meaningful interactions \citep{burke2010social}. News and informational content generating unexpectedly high retweets could be prioritized in discovery feeds for knowledge-seeking users, while content with high comment rates could feature prominently in community-building contexts where discussion adds particular value \citep{bernstein2013quantifying}. Interface design could similarly benefit from these insights through context-aware presentation—highlighting comment counts for political content where debate enriches user experience, while emphasizing sharing metrics for informational content where broad distribution serves the public interest \citep{levordashka2016s}. These targeted approaches could significantly enhance user satisfaction by aligning platform mechanics with the distinct purposes different content types serve.}

\add{Content creators can leverage these findings to develop strategic, objective-specific approaches rather than pursuing generic ``engagement'' goals \citep{meier2014more}. For maximizing information dissemination, creators should focus on objective content with moderate complexity and include relevant URLs, as these features consistently drive unexpected retweets across domains \citep{tan2014effect}. Those aiming to build community through discussion should incorporate personal connection elements while maintaining sufficient depth to warrant thoughtful responses, as these characteristics generate disproportionate comment rates \citep{boyd2010tweet}. For relationship development and brand affinity, emotionally resonant, subjective content tends to generate unexpected likes, suggesting creators should emphasize authentic personal narrative when seeking stronger audience connections \citep{stieglitz2013emotions}. These domain-specific strategies allow creators to move beyond one-size-fits-all approaches to craft content specifically optimized for their communication goals. As social media increasingly shapes public discourse and interpersonal communication, these nuanced insights into engagement drivers offer both theoretical advancements and practical strategies for navigating digital communication landscapes \citep{burke2010social}, with implications for content strategy, user experience design, and analytics tools that better reflect the multifaceted nature of online interactions \citep{bernstein2013quantifying}.}

\section{Related Work}
\add{\subsection{Social Media Engagement Metrics and Patterns:}
Social media engagement research has evolved from examining individual interaction types to considering multiple engagement metrics concurrently \citep{tenenboim2015prompts, aldous2019view, pancer2016popularity}. Engagement on platforms like Twitter follows power-law distributions, where a small fraction of content receives disproportionately high engagement while most receives minimal attention \citep{bakshy2011everyone, coscia2017popularity}. Early studies tended to examine interaction modalities separately, revealing distinct motivations behind favoriting, sharing, and commenting behaviors \citep{levordashka2016s, boyd2010tweet, meier2014more}. However, more recent work has begun exploring the relationships between these metrics, with researchers like \citet{tenenboim2015prompts} demonstrating that clicks, shares, and comments on news content often follow divergent patterns driven by different user motivations. Similarly, \citet{aldous2019view} analyzed engagement across five social media platforms, finding that interaction metrics exhibit platform-specific relationships that reveal underlying user behavior patterns. \citet{pancer2016popularity} further demonstrated that different engagement metrics on political content follow distinct trajectories influenced by message attributes and timing. Despite these advances, most existing research treats engagement primarily as a volume-based metric—focusing on maximizing overall interaction counts rather than understanding unexpected patterns in engagement composition~\citep{scissors2016,burke2010social}. Our work extends this literature by examining scenarios where content receives disproportionate engagement of specific types relative to others, revealing patterns that challenge the conventional treatment of engagement as a uniform, construct.

\subsection{Topic-Based Drivers of Social Media Engagement:}
Content topic consistently emerges as a primary driver of engagement across social media platforms, with domains like music, sports, and news generating distinctive interaction profiles \citep{meier2014more, gilbert2013widespread}. These topic-based variations are not merely differences in volume but reflect fundamentally different diffusion mechanisms and user motivations \citep{romero2011differences, yang2012we}. Sports and political messages, for instance, have been shown to diffuse more widely through networks, often following distinct temporal patterns that reflect their real-world contexts \citep{gonzalez2010structure, yardi2010dynamic}. Political content tends to generate sustained engagement through extended discussions, particularly when it contains controversial viewpoints that spark debate~\citep{gonzalez2010structure,yujiangdhillon2024}. In contrast, sports-related content typically experiences sudden engagement spikes during live events, followed by rapid decay \citep{lehmann2012dynamical}. Entertainment and celebrity content often generates high like-to-share ratios, reflecting passive consumption rather than active propagation \citep{suh2010want, kwak2010twitter}. The role of trending topics in amplifying engagement has been well-documented \citep{boyd2010tweet,lehmann2012dynamical}, indicating that the temporal and cultural context of content significantly influences its spread and engagement patterns. These topic-based variations suggest that users employ different evaluation criteria when deciding how to interact with content from different domains—a phenomenon that our research examines by focusing specifically on unexpected deviations in engagement composition rather than merely overall popularity.

\subsection{Emotional Valence and Content Sentiment Effects:}
The role of content valence in driving engagement patterns has emerged as a crucial factor across social media research, with studies consistently demonstrating that emotional content generates distinctly different interaction profiles compared to neutral information \citep{meier2014more, berger2012makes, stieglitz2013emotions}. This emotional influence operates across multiple dimensions, affecting not only whether users engage but also how they choose to interact with content. Messages evoking high-arousal emotions such as awe, anger, or anxiety tend to spread more rapidly and broadly than content associated with low-arousal emotions like sadness or contentment \citep{stieglitz2013emotions,hutto2014vader,berger2012makes}. Positive content generally receives more likes, while controversial or negative content often generates more comments but fewer endorsements \citep{hansen2011good, naveed2011bad}. Political and social issue discussions demonstrate particularly strong emotional effects, with messages expressing strong positive or negative sentiments propagating more effectively through networks \citep{gonzalez2010structure, tan2014effect}. Research by \citet{berger2012makes} and \citet{brady2017emotion} has revealed complex relationships between specific emotions and virality, showing that content triggering moral outrage or awe achieves broader spread than content eliciting sadness or satisfaction. The emotional component appears to play a crucial role in users' decisions about how to engage with content—whether through simple acknowledgment (likes), amplification (retweets), or discussion (comments). While research has established that emotions drive overall engagement levels, less attention has been paid to how emotional qualities might contribute to unexpected deviations in engagement composition, a gap our study addresses through its ``unexpectedness quotient'' approach.

\subsection{Textual Complexity and Content Accessibility:}
Textual complexity represents another significant factor in engagement patterns, with substantial evidence demonstrating that a post's informational content and ease of comprehension significantly influence how users interact with it \citep{boyd2010tweet, meier2014more, tan2014effect}. The relationship between complexity and engagement follows a non-linear pattern, with moderately complex content—balancing informativeness with accessibility—generally receiving optimal engagement levels \citep{suh2010want, tan2014effect, naveed2011bad}. Several specific content features have been identified as key determinants of engagement: URL inclusion typically increases sharing rates by 30-50\%, as users value content that provides additional context or information depth \citep{suh2010want, tan2014effect}; optimal text length varies by platform, with Twitter's character limitations creating unique dynamics where longer tweets (within platform constraints) often receive higher engagement than very short messages \citep{tan2014effect, boyd2010tweet}; and linguistic concreteness affects sharing patterns, with more concrete language generally driving higher engagement across platforms \citep{berger2012makes}. Content readability also influences engagement type, with more complex vocabulary and sentence structures generating different engagement profiles than simple, accessible language \citep{tan2014effect, naveed2011bad}. Research examining these factors has traditionally focused on maximizing overall engagement rather than understanding how complexity features might drive asymmetric engagement patterns across different interaction types. Our study extends this literature by examining how textual complexity characteristics might systematically predict unexpected deviations in engagement composition—particularly how factors like URL inclusion, content length, and readability might drive disproportionately high likes, retweets, or comments relative to other engagement types.

\subsection{Theoretical Frameworks for Understanding Engagement:}
Two theoretical frameworks provide particularly valuable lenses for understanding social media engagement patterns: Signaling Theory and Attention Economy Theory. Signaling Theory, originally developed to explain communication through observable actions \citep{donath2007signals, connelly2011signaling}, has been increasingly applied to social media contexts to understand how different engagement behaviors function as signals of varying strength and cost. In this framework, likes represent low-cost signals of approval or acknowledgment, while comments require greater cognitive investment and thus serve as stronger signals of engagement \citep{sharma2012inferring, donath2007signals}. Retweets occupy an intermediate position, representing a moderate cost signal that amplifies content beyond the original audience \citep{boyd2010tweet, macskassy2011people}. While previous applications of Signaling Theory tend to treat these engagement types as having fixed relative costs, our work examines how content characteristics might systematically influence signal utilization, revealing more nuanced signaling behaviors than previously recognized. Complementing this approach, Attention Economy Theory conceptualizes attention as a scarce resource that content must compete for in information-rich environments \citep{simon1996designing, franck2019economy, davenport2001attention}. Research applying this framework to social media has typically focused on how content succeeds or fails in capturing user attention overall, rather than examining how different content features might compete for specific types of attention investment \citep{webster2014marketplace, huberman2008social}. By examining how certain content features systematically drive unexpected patterns in engagement composition, our study extends both theoretical frameworks, revealing how users allocate different types of attention investment based on content characteristics and how the perceived cost and value of different engagement signals varies systematically across content domains.}

\section{Empirical Setup}
To investigate the factors contributing to unexpected engagement patterns on Twitter, we develop a robust empirical framework that leverages a large-scale dataset and incorporates a comprehensive set of content and author features. This section details our data collection process, the criteria used for tweet selection, and the feature generation methods employed in our analysis.
\subsection{Data Description}

\del{Our study utilizes a large dataset derived from the Twitter Decahose, which provides a 10\% random sample of all Twitter data. We focused our analysis on English-language tweets posted between January 1, 2018, and December 31, 2018, offering a full year of data to capture potential seasonal variations in engagement patterns. We used the data for last uninterrupted year prior to the start of the COVID-19 pandemic in December 2019. To ensure our study captures the most relevant and impactful content, we constrained our analysis to tweets containing one or more of the top 500 hashtags used in the Decahose during 2018, using these hashtags as proxies for tweet topics.}

\add{Our study utilizes a large dataset derived from the Twitter Decahose, which provides a 10\% random sample of all Twitter data. We focused on English-language tweets posted between January 1, 2018, and December 31, 2018, capturing a full year of potential seasonal variations in engagement patterns \citep{golder2014digital, jungherr2015analyzing}. We selected tweets containing one or more of the top 500 hashtags used in the Decahose during this period to facilitate reliable topic identification \citep{yang2012we, posch2013meaning}. Recognizing the potential selection bias in focusing on hashtag-containing tweets, we conducted a comparative analysis with non-hashtag tweets using a supplementary random sample. While hashtag tweets received moderately higher overall engagement (approximately 15\% more on average), the relationships between content features and engagement types remained consistent across both groups.}

We excluded tweets lacking textual content, as our primary aim is to unpack the textual drivers of social media engagement. To facilitate our investigation into the differential impact of various interaction modalities, we only included tweets that received at least one like, one retweet, and one comment—ensuring that all engagement types were represented in our analysis. After applying these filtering criteria, our final dataset comprises 642,108 unique tweets, providing a robust sample for drawing substantiated conclusions about the factors influencing unexpected engagement patterns on Twitter. This comprehensive dataset, collected during a period of relative platform stability, offers a particularly valuable window into enduring content-engagement relationships.

\subsubsection{Dataset Rationale and Temporal Context:} 
\add{Our choice of 2018 data was deliberate for several methodological and practical reasons. First, 2018 represents the last complete year before two significant disruptions that fundamentally altered social media dynamics: the COVID-19 pandemic, which dramatically shifted online behavior patterns in 2019, and Twitter's subsequent rebranding as X with accompanying API access restrictions that have made comprehensive academic research increasingly challenging \citep{freelon2018computational,bruns2021after}. This timing provides a valuable baseline for understanding fundamental content-engagement relationships absent these extraordinary influences. The continued relevance of our findings also is supported by recent comparative studies. \citet{shahbaznezhad2021role} examined social media engagement patterns across multiple time periods and found that while absolute engagement levels fluctuate over time, the relative relationships between content features and engagement types remain relatively consistent. Similarly, \citet{moran2020message} demonstrated that content characteristics that drove specific engagement behaviors in 2018-2019 continued to show similar relationships in 2020-2021, despite interface changes and shifting usage patterns.

Our focus on hashtag-containing tweets provides methodological advantages for reliable topic identification, following established approaches in computational social media research \citep{yang2012we, wang2016using, posch2013meaning}. Hashtags serve as explicit topic markers that allow for more accurate content categorization compared to automated topic extraction methods, which often struggle with Twitter's short-form content \citep{yang2012we}. This pre-pandemic dataset also provides a cleaner baseline for understanding content-engagement relationships without the confounding effects of the extraordinary global events that significantly altered social media usage patterns across all platforms. By establishing these relationships in a relatively stable period, our findings can serve as a reference point for future studies examining how these dynamics might have shifted during and after major disruptions to online communication norms. The patterns we identify likely reflect fundamental aspects of user behavior that transcend specific platform iterations.}

\subsection{Feature Generation}
To comprehensively analyze the factors contributing to unexpected engagement, we generated two primary categories of features: content features and author features. These feature sets allow us to examine both the characteristics of the tweets themselves and the attributes of the users who created them, providing a holistic view of the elements that may influence engagement patterns.

\subsubsection{Textual complexity features}

First and foremost, we construct a set of features representing the complexity of the tweet. The textual complexity measures how well a reader can process the tweet's content. Users are likely to engage more with content that they can easily comprehend and relate to. We consider four key aspects of textual complexity:

\begin{itemize}
    \item {\it Readability:} We derive our readability feature based on the Dale-Chall readability formula \citep{kincaid1975derivation}. The Dale-Chall formula is a readability test that quantifies a text's comprehension difficulty. A higher score indicates that the text is difficult to grasp, and lower scores denote material that is easy to read and understand. This measure allows us to assess how the complexity of language used in a tweet might influence its engagement patterns.
   
    \item {\it Concreteness:} This feature quantifies the concreteness of the tweets. We code this feature using the concreteness rating of 40,000 common English word lemmas annotated by \citet{brysbaert2014concreteness}. This lexicon rates the concreteness of each word on a scale of [1-5], with 1 indicating a highly abstract word and 5 denoting a very concrete word. Finally, the concreteness of a tweet is calculated as the average of the concreteness of all its constituent words. By measuring concreteness, we can explore how the use of abstract versus concrete language affects user engagement.
    
    \item {\it Length:} This feature encodes the number of characters in the tweet after removing links and hashtags. For example, the post {\it Happy Birthday :)} has a length of 16, including spaces and the smiley. Tweet length serves as a proxy for the amount of information contained in a tweet and allows us to investigate how the volume of content influences engagement patterns.
  
    \item {\it Link:} It is a binary feature that indicates if the tweet includes an external URL link. Our dataset contains 495,646 tweets with links and 146,462 tweets without links. This feature enables us to examine how the inclusion of external resources affects user engagement, potentially reflecting the tweet's informational value or its role in directing user attention beyond the platform.
\end{itemize}

\begin{figure}[htbp]
\centering
\includegraphics[width=0.45\textwidth, center]{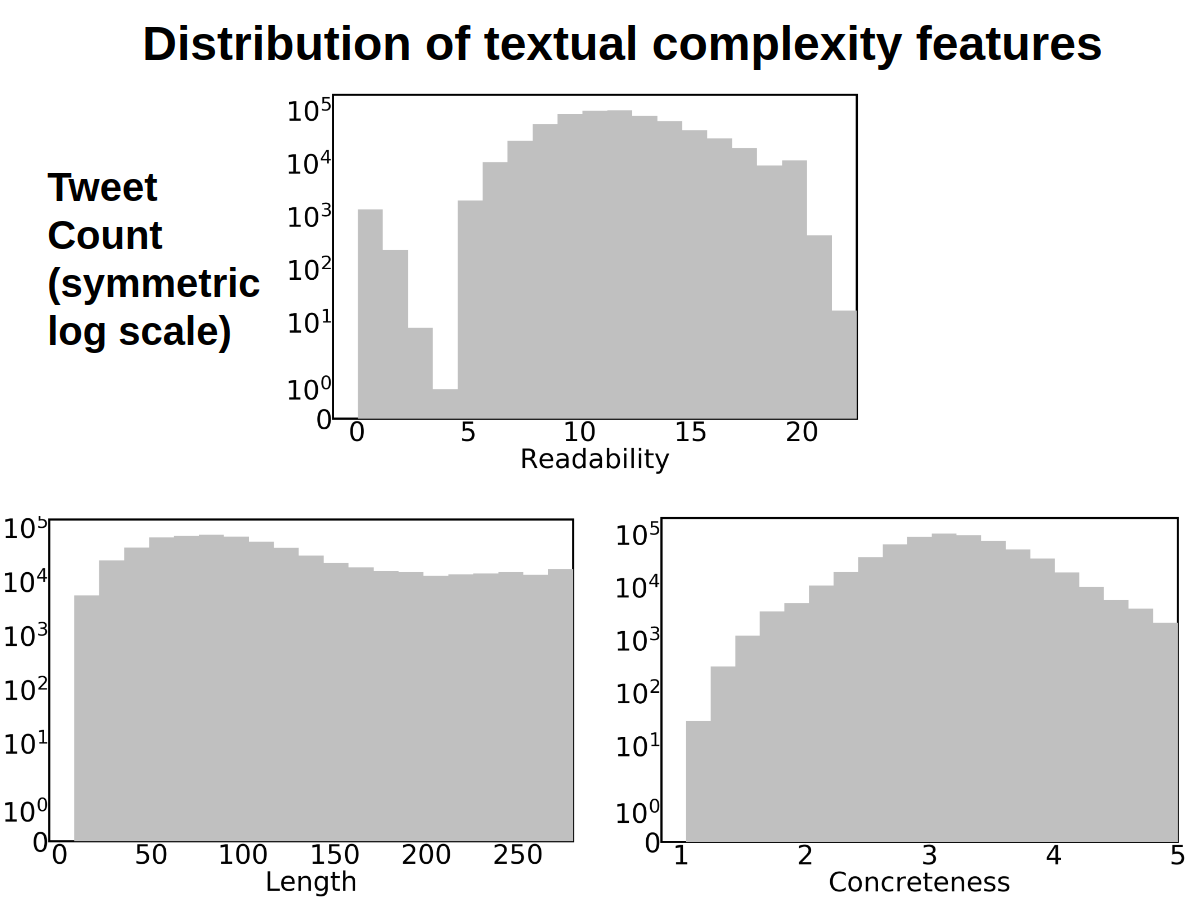}
\caption{Empirical distribution of the readability, length, and concreteness features.
} 
\label{fig:textualcomp}
\end{figure}

Figure \ref{fig:textualcomp} shows the empirical distribution of the various textual complexity features in our dataset, providing a visual representation of the diversity in textual complexity across the tweets we analyzed.

\subsubsection{Valence features}

The affective quality of tweets, particularly the emotions expressed in the content, is a crucial driver of engagement on social media platforms. To capture these emotional aspects, we employ two key valence features that allow us to quantify the sentiment and subjectivity of each tweet. These features provide insights into how emotional content and personal opinions influence user engagement patterns:

\begin{itemize}
    \item {\it Sentiment:} To measure the overall emotional tone of each tweet, we utilize the widely-recognized VADER (Valence Aware Dictionary and sEntiment Reasoner) algorithm \citep{hutto2014vader}. This sophisticated sentiment analysis tool is specifically attuned to sentiments expressed in social media contexts. VADER computes a sentiment score on a continuous scale from -1 to +1, where -1 represents the most negative sentiment, +1 indicates the most positive sentiment, and 0 signifies a neutral tone. This nuanced approach allows us to capture the full spectrum of emotional expression in tweets, from strongly negative to strongly positive, and examine how varying degrees of sentiment correlate with different types of user engagement.

    \item {\it Subjectivity:} To complement our sentiment analysis, we also measure the subjectivity of each tweet. This feature quantifies the degree to which the tweet expresses personal opinions or emotions versus objective facts or information. We code subjectivity on a scale from 0 to 1, where 0 represents highly objective content and 1 denotes very subjective content. To calculate this measure, we employ the method developed by \citet{pang2004sentimental}, implemented through the TextBlob package\footnote{\url{https://planspace.org/20150607-textblob_sentiment/}}. By including this feature, we aim to investigate whether there are significant differences in the types and levels of engagement attracted by subjective personal opinions compared to more objective, factual content. This analysis is particularly relevant on Twitter, where both personal views and factual information are known to drive high levels of engagement.
\end{itemize}

\begin{figure}[htbp]
\centering
\includegraphics[width=0.45\textwidth]{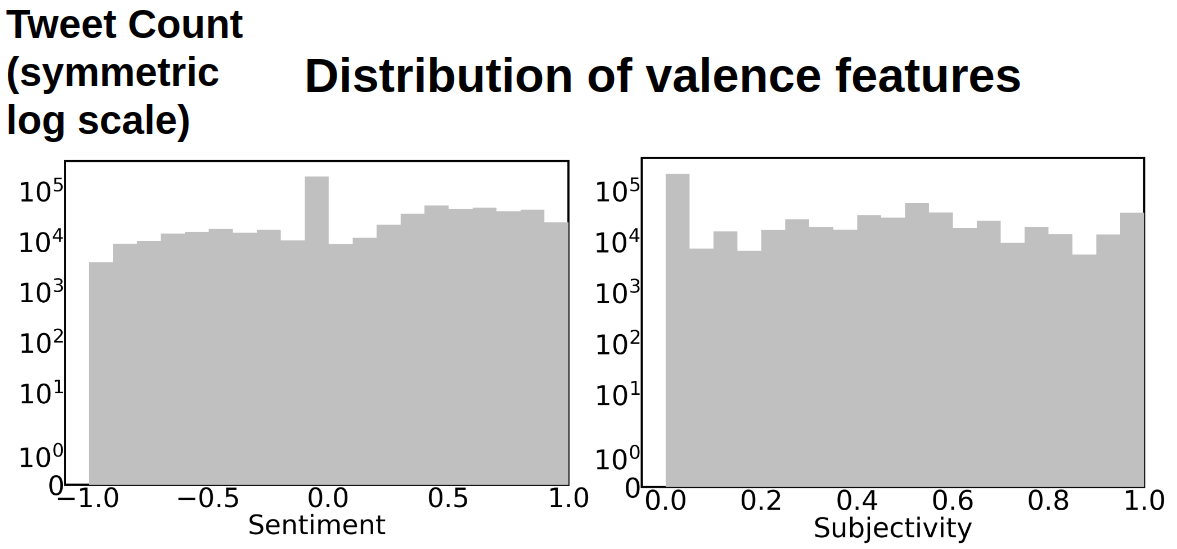}
\caption{Empirical distribution of the sentiment and subjectivity features.
}
\label{fig:valenceFeatures}
\end{figure}

Figure \ref{fig:valenceFeatures} presents the empirical distribution of both valence features across our dataset. These distributions provide valuable insights into the emotional landscape of our tweet corpus, showcasing the range of sentiment and subjectivity present in the content we analyzed. By examining these distributions in conjunction with engagement patterns, we can uncover important relationships between the emotional content of tweets and the types of user interactions they elicit.

\subsubsection{Topic Features}

The topic of a tweet is undeniably a crucial driver of its engagement, playing a significant role in determining how users interact with the content. To systematically identify and categorize the topical composition of tweets in our dataset, we have developed a comprehensive approach that builds upon and extends the methodology of \citet{romero2011differences}, who annotated tweet hashtags with eight distinct topic labels. Recognizing the dynamic nature of social media discourse and potential shifts in popular topics over time, we expanded this classification system by adding three new categories: Business, News, and Arts. This expansion was based on a meticulous manual inspection of the top 500 hashtags in our dataset and analysis of random samples of their corresponding tweets. Table \ref{tab:topic_definitions} provides detailed descriptions of these new topic categories, elucidating their scope and significance within our classification system.

Using this enhanced set of 11 topic labels, we annotated the top 500 hashtags in our dataset, allowing for multi-label classification to acknowledge that many hashtags can span multiple thematic areas. For instance, the hashtag \#cryptocurrency is categorized under both Business and Technology. Hashtags that did not align with any of the 11 defined topic categories were classified as "Other" to ensure comprehensive coverage. We then applied these annotated hashtags to identify the topical composition of individual tweets, noting that this method allows for tweets to be associated with multiple topics, mirroring the multifaceted nature of many social media posts.

Table \ref{tab:topic_cnt_example} presents a detailed breakdown of our topic classification results, including the categories of hashtags, the number of tweets associated with each topic, and representative examples. This table not only illustrates the distribution of topics across our dataset but also provides concrete examples that demonstrate the practical application of our classification system. By employing this sophisticated topic classification methodology, we aim to provide a robust foundation for analyzing how different subject matters influence engagement patterns on Twitter, potentially uncovering valuable insights into what types of topics are most likely to generate unexpected engagement across various interaction modalities.

\begin{table}[htbp]
  \caption{Defining the scope of the Business, News, and Arts topics.}
  \label{tab:topic_definitions}
  \begin{tabular}{l p{6.4cm}}
    \toprule
    \textbf{Topics} & \textbf{Definition}\\
    \midrule
    Business & Names of financial instruments or financial products and concepts related with entrepreneurship, marketing, accounting, finance, and investment.\\
    
    News  & Names of news platforms and hashtags indicating that the tweet contains news content.\\
    
    Arts  & Hashtags related to visual arts, design, fashion, dance, architecture, and poetry.\\
  \bottomrule
    \end{tabular}
\end{table}

\begin{table}[htbp]
  \caption{Examples of hashtags belonging to each topic. We extend the hashtags described in~\cite{romero2011differences} with Art, Business, and News topics.}
  \label{tab:topic_cnt_example}
  \begin{tabular}{lcp{4cm}}
        \toprule
        \textbf{Topics} & \textbf{\# Tweets} & \textbf{Sample hashtags} \\
        \midrule
        Technology &  { 41,932} & \#machinelearning, \#robotics \\
        
        Sports &  { 57,890}   & \#worldcup, \#football \\
        
        Games  & { 32,210}   &  \#ps4live, \#fortnite   \\
        
        Idioms &  { 59,998}  & \#nowplaying, \#fridayfeeling  \\
        
        Movies/TV &  { 89,771 } &  \#loveisland, \#bigolive  \\
        
        Politics & { 74,714 } & \#democrats, \#trump \\
        
        Music  & { 96,254  } & \#newmusic, \#soundcloud \\
        
        Celebrity  & { 60,048} & \#bts, \#selenagomez \\
        
        Art  &  { 27,408}  & \#design, \#photography \\
        
        Business  &  { 30,768}  & \#marketing, \#forex \\
        
        News  &  { 17,212}  & \#foxnews, \#breakingnews \\
        
         Other  & { 202,467}  & \#china, \#school \\
        \bottomrule     
    \end{tabular}
\end{table}

\subsubsection{Author attributes}
While this paper primarily focuses on assessing the impact of a tweet's content attributes on its engagement, it is crucial to acknowledge that the characteristics of the tweet's author play a significant role in determining its popularity and reach \citep{bakshy2011everyone,cha2010measuring}. To ensure a comprehensive and accurate analysis, we incorporate several key author attributes as control variables in our study. Drawing from established research methodologies \citep{golder2014digital, jungherr2015analyzing} we identify and account for three primary author features that have been shown to influence tweet engagement patterns: the number of followers, membership on public lists, and verification status \citep{bakshy2011everyone,paul2019elites}.

\begin{figure}[ht]
\centering
\includegraphics[width=0.45\textwidth]{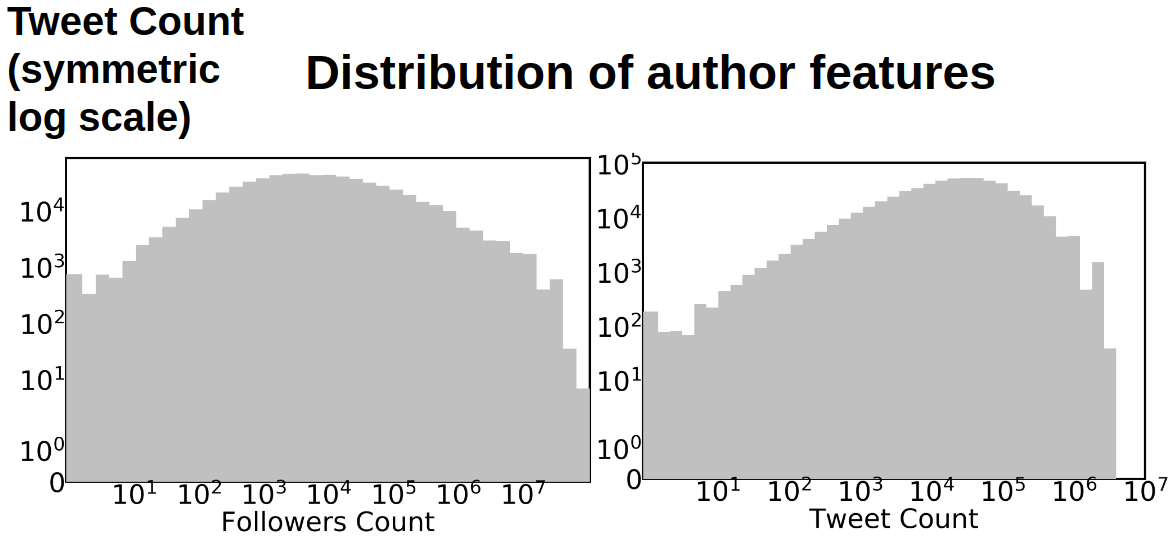}
\caption{Empirical distribution of the author features: number of followers and number of tweets.}

\label{fig:authorfeatures}
\end{figure}

\begin{itemize}
    \item {\it Followers:} The number of followers an author has is a fundamental indicator of their potential reach and influence on the platform. We control for this variable to account for the varying levels of visibility different authors may have. Figure \ref{fig:authorfeatures} illustrates the empirical distribution of follower counts across our dataset, revealing the diverse range of author popularities represented in our study. This distribution allows us to consider how the size of an author's audience might interact with content features to affect engagement patterns.

    \item {\it Lists:} Another metric of author popularity and influence is the number of public lists that include the tweet author. This feature provides insight into the author's perceived value within specific communities or interest groups on Twitter. By controlling for list memberships, we can account for the potential impact of an author's reputation or expertise in particular domains on tweet engagement.

    \item {\it Verified:} Twitter's verification system bestows a unique form of social authority on certain accounts, which has been demonstrated to influence engagement patterns. We include a binary feature indicating whether the author has a verified Twitter account. In our dataset of 642,108 tweets, 112,009 were authored by verified users. This substantial proportion of verified content allows us to examine how this official recognition interacts with content features to affect unexpected engagement patterns.
\end{itemize}

By incorporating these author attributes into our analysis, we aim to isolate the effects of content features on unexpected engagement patterns more accurately. This approach allows us to disentangle the influence of an author's established popularity or authority from the intrinsic characteristics of the tweet content itself, providing a more nuanced understanding of what drives unexpected engagement on Twitter.

\subsection{Twitter Engagement Types}

Twitter, widely recognized as both a social network and a news medium \citep{kwak2010twitter}, offers its users a range of active interaction options with content. These engagement types not only reflect different levels of user involvement but also convey distinct social signals and require varying degrees of effort. Understanding these interaction modalities is crucial for our analysis of unexpected engagement patterns. The platform primarily facilitates three main types of user engagement:

\begin{itemize}
    \item {\it Like/Favorite:} This interaction type represents the lowest barrier to engagement, requiring only a single click from the user. Likes or favorites often signal support, agreement, or enthusiasm for the content. The ease of this action makes it a common form of engagement, but its ubiquity also means that an unexpectedly high number of likes may be particularly significant. In the context of Signaling Theory, likes can be viewed as a low-cost signal of approval or acknowledgment.

    \item {\it Retweet/Share:} Retweeting or sharing a tweet represents a higher level of engagement compared to liking. When a user retweets content, they are choosing to disseminate that information to their own follower network, implicitly endorsing its value or relevance. However, it's important to note that retweets do not always signal agreement or endorsement; they may also be used to spread information neutrally or even critically. In terms of the Attention Economy, retweets play a crucial role in amplifying content and extending its reach beyond the original author's network.

    \item {\it Comment:} Commenting on a tweet represents the highest form of engagement among these three types. It requires active participation in the conversation, demonstrating a willingness to contribute one's own thoughts or opinions. Comments not only demand more time and cognitive effort from users but also expose them to potential responses from others, making this a more high-stakes form of interaction. From the perspective of Signaling Theory, comments can be seen as a strong signal of interest or investment in the topic.
\end{itemize}

\begin{figure}[ht]
\centering
\includegraphics[width=0.45\textwidth]{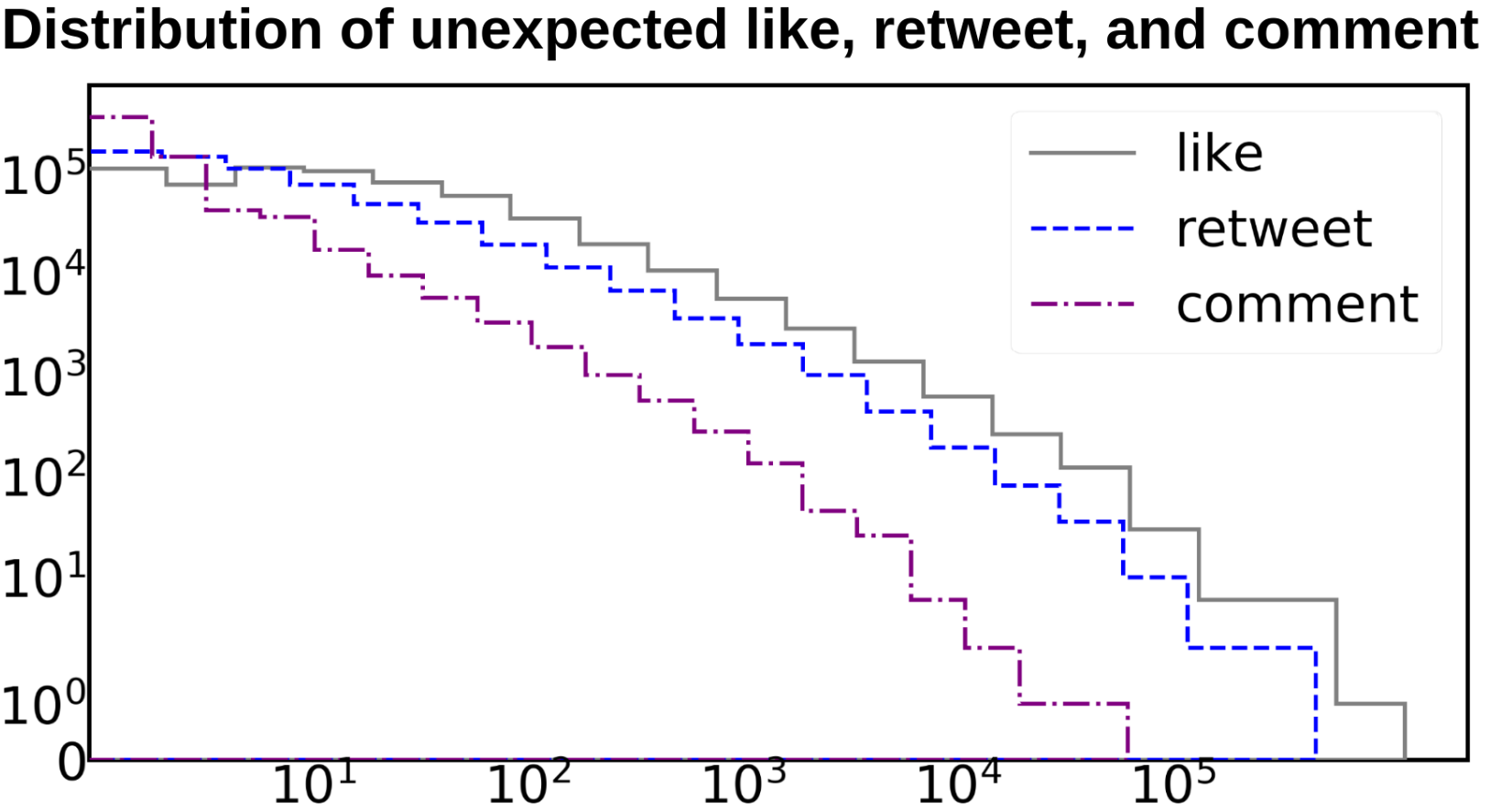}
\caption{Empirical distribution of the number of likes, retweets, and comments.}
\label{fig:summarydist}
\end{figure}

Figure \ref{fig:summarydist} provides a visual representation of the empirical distribution of likes, retweets, and comments received by the tweets in our dataset. This distribution offers valuable insights into the relative frequency and scale of each engagement type, setting the stage for our analysis of unexpected engagement patterns. It's worth noting that while our study focuses on Twitter, similar forms of active interactions are common across other social media platforms such as Facebook and Reddit, potentially allowing for broader applicability of our findings. The varying levels of effort and signaling strength associated with each engagement type provide a rich framework for examining how different content features might elicit unexpected patterns of interaction, contributing to our understanding of user behavior in the attention economy of social media.

\section{Empirical Models \& Results}
Our empirical analyses proceed in two steps, designed to uncover the nuanced dynamics of social media engagement through the lenses of Signaling Theory and Attention Economy Theory. First, we develop a novel metric to capture unexpectedly high engagement with a tweet, addressing a crucial gap in our understanding of how content breaks through the noise in the attention economy of social media. Second, we conduct a regression analysis to identify the content features that contribute to higher than predicted engagement, using our newly defined unexpectedness measure as the dependent variable. This approach allows us to explore how different content characteristics act as signals and compete for user attention in ways that deviate from typical patterns.

\subsection{Step 1: Defining the "unexpectedness" quotient}
In the complex ecosystem of social media engagement, defining what constitutes unexpectedly high engagement presents a unique challenge. This challenge arises from the inherent correlations between different types of engagement - likes, retweets, and comments - as illustrated in Table \ref{corr2}. These strong correlations necessitate a more nuanced approach to identifying truly unexpected engagement patterns.

\begin{table}[htbp]
\centering
\small
\begin{tabular}{l|l|l}
\hline
\hline
         & $\text{N}^{\text{Like}}_i$                                           & $\text{N}^{\text{Retweet}}_i$                    \\
\hline
$\text{N}^{\text{Retweet}}_i$    & 0.96    &      -                    \\
\hline
$\text{N}^{\text{Comment}}_i$ & 0.80 & 0.81   \\
\end{tabular}
\caption{Correlation between the number of like, retweet, and comment in our dateset}
\label{corr2}
\end{table}

The concept of unexpectedness is crucial in the context of both Signaling Theory and Attention Economy Theory, as it helps identify content that breaks through typical user interaction patterns. In Signaling Theory, unexpected signals are particularly potent, as they stand out from the background noise and potentially convey more information about user behavior and content value. In the attention economy of social media, content that generates unexpected engagement patterns may be seen as capturing a disproportionate share of the limited resource of user attention. Our unexpectedness quotient is designed to quantify these phenomena by measuring deviations from expected engagement, given all other engagement types. This approach allows us to control for the overall popularity of a tweet and focus on the specific ways in which it may outperform expectations for a particular type of engagement. For instance, we can use a robust model to predict a tweet's number of likes based on its retweets and comments, and then define the unexpectedness of the tweet's likes as the deviation from this predicted value. We apply this methodology consistently across all three engagement types - likes, retweets, and comments, enabling a comprehensive analysis of unexpected engagement patterns.

A key methodological challenge in identifying unexpected engagement patterns lies in the complex nature of social media interaction data. Social media engagements typically follow a power-law distribution, where a small number of tweets receive extremely high engagement while the majority receive moderate to low engagement \citep{bakshy2011everyone, coscia2017popularity}. This characteristic makes traditional mean-based regression methods inadequate for our analysis, as they are sensitive to outliers and assume normally distributed data. Moreover, the three types of engagement—likes, retweets, and comments—show strong correlations with each other, making it crucial to develop a method that can account for these relationships while identifying truly unexpected patterns. Additional complexity arises from the heterogeneous nature of engagement across different content types and user communities. The skewed nature of the data, combined with these intricate relationships between engagement types, necessitates a sophisticated analytical approach that can handle non-normal distributions while maintaining interpretability. These challenges are further compounded by the temporal dynamics of social media engagement, where the relationship between different types of engagement may vary over time.

To address these methodological challenges, we employ quantile regression, a robust statistical technique that offers several key advantages for our analysis. Unlike ordinary least squares (OLS) regression, which minimizes the sum of squared residuals to estimate the conditional mean, quantile regression minimizes the sum of asymmetrically weighted absolute residuals to estimate the conditional quantiles. This approach allows us to model relationships between variables at different points of the response distribution rather than just at the mean, making it particularly suitable for our skewed social media data. The method's robustness to outliers and ability to capture heterogeneous effects across different levels of engagement enables us to understand how relationships between engagement types might vary across the distribution. We specifically estimate the 90th percentile quantile regression (though our results remain robust when using the 75th and 50th percentiles, as discussed later), as described in Equations \ref{predict:like}, \ref{predict:rt}, and \ref{predict:c}, to predict a tweet's expected engagement metrics. The choice of the 90th percentile helps establish a meaningful benchmark for identifying truly exceptional engagement patterns while maintaining statistical robustness. This quantile regression framework provides a solid foundation for developing our unexpectedness quotient, allowing us to account for the complex relationships between different types of engagement while identifying meaningful deviations from expected patterns.

\begin{eqnarray}
\label{predict:like}
&\text{Likes}_i \sim  \text{Quantile}(\text{Retweets}_i + \text{Comments}_i)\\
\label{predict:rt}
&\text{Retweets}_i \sim  \text{Quantile}(\text{Likes}_i + \text{Comments}_i)\\ 
\label{predict:c}
&\text{Comments}_i \sim  \text{Quantile}(\text{Likes}_i + \text{Retweets}_i)
\end{eqnarray}

Building on these regression models, we develop an unexpectedness quotient that provides an intuitive yet powerful measure of engagement deviation. This metric, defined as the ratio of observed to predicted engagement, quantifies how much a tweet's actual engagement deviates from what would be expected given its performance across other engagement types. The quotients, shown in Equations \ref{unexp:like}, \ref{unexp:rt}, and \ref{unexp:c}, offer several advantages for analyzing engagement patterns. First, they provide a normalized measure that allows for comparison across tweets with different absolute levels of engagement, enabling meaningful analysis across diverse content types and user communities. Second, by using the 90th percentile predictions as the denominator, we establish a robust benchmark for ``expected'' engagement levels that accounts for the natural skewness in social media data. Third, this approach allows us to identify truly exceptional cases where one type of engagement significantly exceeds what would be predicted based on the other types, helping us isolate and study these unexpected patterns. The quotient's design also maintains interpretability while capturing the complex relationships between different forms of engagement, making it a valuable tool for both researchers and practitioners interested in understanding social media dynamics. Moreover, the metric's flexibility allows it to be applied across different platforms and contexts, potentially offering insights into broader patterns of online social interaction.

\begin{eqnarray}
\label{unexp:like}
E_i^{\text{Likes}}&=& \frac{\text{Likes}_i}{\widehat{\text{Likes}}_i}\\
\label{unexp:rt}
E_i^{\text{Retweets}}&=& \frac{\text{Retweets}_i}{\widehat{\text{Retweets}}_i}\\
\label{unexp:c}
E_i^{\text{Comments}}&=& \frac{\text{Comments}_i}{\widehat{\text{Comments}}_i}
\end{eqnarray}

The unexpectedness quotient serves as a bridge between our empirical observations and the theoretical frameworks of Signaling Theory and Attention Economy Theory. In terms of Signaling Theory, a high unexpectedness quotient for a particular engagement type might indicate that the content is sending a particularly strong or unusual signal to users, prompting them to engage in ways that deviate from typical patterns. For instance, a tweet with an unexpectedly high number of comments might be seen as prompting a stronger signal of engagement, as commenting requires more effort and cognitive investment than liking or retweeting.

From the perspective of Attention Economy Theory, content with high unexpectedness quotients can be viewed as particularly successful in capturing user attention in a competitive information environment. These tweets have managed to not only engage users but to do so in ways that exceed the typical relationships between different forms of engagement. This could indicate content that is especially compelling, controversial, or relevant to users' interests, allowing it to stand out in the crowded social media landscape.

\begin{figure}[htbp]
\tiny
\centering
\includegraphics[width=0.5\textwidth]{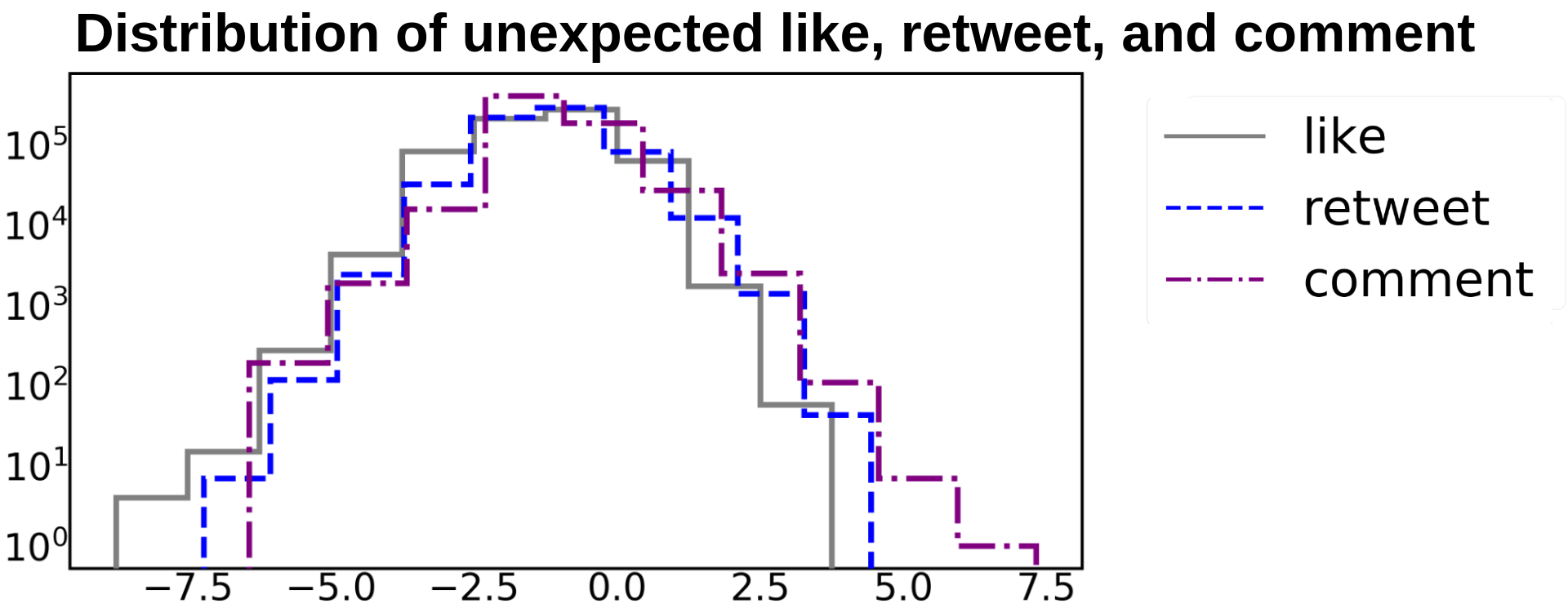}
\caption{Unexpectedness quotient for likes, retweets, and comments.}
\label{fig:unexpected}
\end{figure}

Figure \ref{fig:unexpected} visualizes the distribution of these unexpectedness quotients across our dataset. Notably, the distribution for comments shows a higher degree of unexpectedness compared to likes and retweets, with these differences proving statistically significant in pairwise t-tests. This finding aligns with both Signaling Theory and Attention Economy Theory. Comments, requiring the most effort from users, can be seen as the strongest signal of engagement. The higher unexpectedness in comments suggests that certain content is particularly effective at prompting this high-effort form of engagement, potentially indicating a more successful capture of user attention and cognitive resources.

To illustrate the practical implications of our unexpectedness quotient, Table \ref{tab:tweet_examples} presents examples of tweets that received unexpectedly high numbers of likes, retweets, and comments. These examples reveal distinct patterns in how different types of content generate unexpected engagement. Tweets receiving unexpectedly high likes typically involve personal achievements, lifestyle content, and emotional expressions, often accompanied by visual content like family photos or selfies. For instance, a tweet sharing a family vacation photo received 814 likes but only 6 retweets, demonstrating how personal content can generate disproportionate expressions of support without broader sharing. In contrast, tweets with unexpectedly high retweets often involve calls to action, such as giveaway promotions or requests for help (e.g., finding lost pets), where sharing serves a clear functional purpose. Perhaps most intriguingly, tweets generating unexpectedly high comments tend to either explicitly solicit responses (e.g., "Write a story for this photo") or address controversial topics that spark discussion (e.g., school safety). These patterns suggest that unexpected engagement often aligns with specific content purposes: emotional connection for likes, information spread for retweets, and community interaction for comments.

By developing and applying this unexpectedness quotient, we provide a novel tool for understanding the dynamics of social media engagement. This metric allows us to move beyond simple counts of engagement and explore the nuanced ways in which content can exceed expectations and capture user attention. The validity of our approach is reinforced by robustness checks showing that our results remain broadly similar when using less extreme quantiles (75\% and 50\%) instead of the 90\% threshold, suggesting that our unexpectedness quotient captures meaningful patterns of engagement across different thresholds of unexpectedness. The metric's ability to identify distinct patterns of unexpected engagement, as demonstrated by the examples in Table \ref{tab:tweet_examples}, suggests its utility for both researchers studying social media dynamics and practitioners seeking to optimize content strategy.

In the following section, we employ this metric to systematically investigate which content features are most associated with unexpected engagement, illuminating the mechanisms by which certain tweets manage to stand out in the crowded social media landscape. By examining how different content characteristics correlate with various types of unexpected engagement, we can better understand the complex interplay between content design, user behavior, and platform dynamics. This analysis not only advances our theoretical understanding of social media engagement but also offers practical insights for content creators and platform designers seeking to facilitate more meaningful online interactions.

\begin{table*}[t]
\centering
\scriptsize                           
\resizebox{\textwidth}{!}{%
\begin{tabular}{p{1.8cm}p{7cm}p{2.2cm}r r r}
\toprule
\textbf{Interaction Type} & \textbf{Example} & \textbf{Summary} & \textbf{Like} & \textbf{RT} & \textbf{Cmt} \\ \midrule

\multirow{5}{*}{\parbox{1.8cm}{\raggedright
Unexpectedly high likes}}
 & (1) Don’t chase waterfalls, chase joy. [Family photo] & \multirow{5}{*}{\parbox{2.2cm}{\raggedright
Typical examples include personal achievements, excitement, and lifestyle updates.}} & 1005 &   9 &   9 \\
 & (2) Brazil, good morning! [Selfie] & & 13008 & 421 & 122 \\
 & (3) My family and me travelling in Campania, Italy. [Photo] & &  814 &   6 &   2 \\
 & (4) We started drinking at 7 PM and now it’s 4 AM. [Selfie] & & 1329 &   7 &  11 \\
 & (5) I’ve bought a beautiful scarf and carpets in Delhi. [Photo] & &  413 &   5 &   2 \\ \addlinespace

\multirow{5}{*}{\parbox{1.8cm}{\raggedright
Unexpectedly high retweets}}
 & (1) Visit our giveaway page to win a gaming chair. & \multirow{5}{*}{\parbox{2.2cm}{\raggedright
Typical examples include “retweet to vote”, lost-and-found notices, and giveaway information.}} &   98 & 1790 &  19 \\
 & (2) Vote for XXX by retweeting this tweet. [Singer photo] & & 1345 & 7433 & 105 \\
 & (3) If this tweet gets 5 k RTs, I’ll rap “Cipher 4” for a month! & & 3855 & 9633 & 251 \\
 & (4) Help me find a lost dog! Here’s her photo. & &   15 &  120 &   1 \\
 & (5) Pretty résumé design for only \$5. Check the link. & &    4 &  140 &   1 \\ \addlinespace

\multirow{5}{*}{\parbox{1.8cm}{\raggedright
Unexpectedly high comments}}
 & (1) Write a story for this photo in six words. [Frog pic] & \multirow{5}{*}{\parbox{2.2cm}{\raggedright
Typical examples include explicit requests for replies and controversial topics.}} &  923 &  205 & 1150 \\
 & (2) If kids are afraid of being shot at school, what should we tell them? & &  142 &   51 &  722 \\
 & (3) Choose my next profile picture: reply 1–4. [Four selfies] & &  625 &  105 & 1422 \\
 & (4) Want more votes for XXX. Reply with \#XXX to enter a prize draw! & &   65 &  178 & 1410 \\
 & (5) Which cryptocurrency will be the next ETH? & &  161 &    6 &  323 \\

\bottomrule
\end{tabular}}%
\caption{Examples of tweets (paraphrased) with unexpectedly high likes, retweets, or comments.}
\label{tab:tweet_examples}
\end{table*}

\begin{figure*}[t]
\centering
\includegraphics[width=0.9\textwidth]{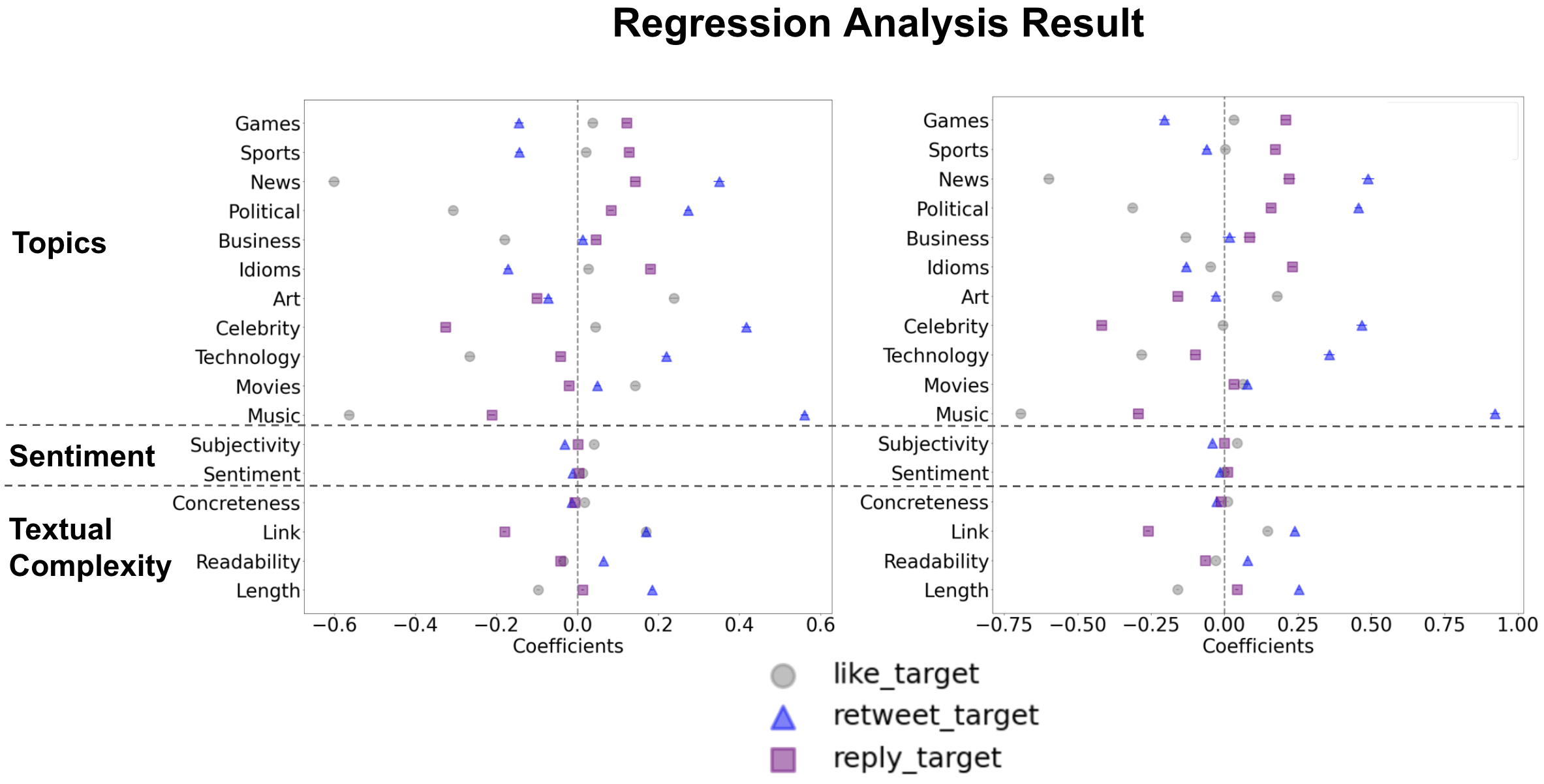}
\caption{OLS Regression estimation results.The left panel (a) shows point estimates for equation  (\ref{predict:OLS-like}) (\ref{predict:OLS-rt}) and (\ref{predict:OLS-c}) using dataset with tweets with at least one like, retweet, and comment. The right panel (b)  shows point estimates  for Equations (\ref{predict:OLS-like}) (\ref{predict:OLS-rt}) and (\ref{predict:OLS-c}) using dataset with tweets with at least one like, retweet, and comment with at least 10 counts in one of the interactions. {\it Note:} 1) Heteroskedasticity Robust standard errors are shown in parenthesis next to the coefficient.  2) The coefficients of the author features are omitted. 3) Error bar indicates 95\% confidential interval. 4) Various statistics for (a) Target variable as unexpected like: log-lik.:  -8.2518e+05 , N: 642108, $R^2$:  0.212, F-statistic:  8218.(a) Target variable as unexpected retweet: log-lik.:  -8.0216e+05, N: 642108, $R^2$:  0.165, F-statistic: 6046 . (a) Target variable as unexpected comment: log-lik.:  -6.9496e+05 N:  642108, $R^2$:  0.090, F-statistic: 3040.0. (b) Target variable as unexpected like: log-lik.: -4.0501e+05, N:  346587, $R^2$:  0.217, F-statistic:  4584.(b) Target variable as unexpected retweet: log-lik.:  -4.6988e+05, N:  346587, $R^2$:  0.235, F-statistic: 5056. (b) Target variable as unexpected comment: log-lik.:  -4.4496e+05, N:  346587, $R^2$:  0.125, F-statistic: 2365.}
\label{fig:regressionResult}
\end{figure*}

\subsection{Step 2: Deconstructing the determiners of unexpected Tweet engagement}
Having established a robust metric for quantifying unexpectedly high engagement with tweets, we now turn our attention to unraveling the key drivers that contribute to these exceptional engagement patterns. Our focus in this step is to deconstruct the impact of content features, particularly textual complexity, valence, and content topic, on driving unexpectedly high engagement. This analysis allows us to explore how different content characteristics function as signals in the attention economy of social media, potentially explaining why certain tweets capture disproportionate user attention and engagement. By examining these factors through the lenses of Signaling Theory and Attention Economy Theory, we can gain deeper insights into the mechanisms that allow certain content to stand out in the crowded digital landscape.

To approach this inquiry, we employ a series of Ordinary Least Squares (OLS) regression analyses, as described in Equations \ref{predict:OLS-like}, \ref{predict:OLS-rt}, and \ref{predict:OLS-c}. In these models, our dependent variable is the log-transformed value of the unexpectedness quotient for likes, retweets, and comments, respectively. We opt for OLS regression due to its simplicity and the ease of interpretability of results, which is particularly valuable when examining the nuanced relationships between content features and engagement patterns. While more complex non-linear models, such as neural networks, might capture additional subtleties in the data, they often sacrifice interpretability, which is crucial for our goal of understanding the specific content attributes that drive unexpected engagement. To address the skewed nature of several of our variables, which can be problematic for OLS models assuming a Gaussian distribution of the dependent variable \citep{angrist2008mostly}, we apply a log-transformation to these skewed variables. This common solution helps to normalize the distribution and improve the model's fit.
\begin{equation}
\label{predict:OLS-like}
E_i^{\text{Likes}} \sim  \text{OLS}(\text{Complexity}_i + \text{Valence}_i+ \text{Topic}_i+ \text{Author}_i)
\end{equation}
\begin{equation}
\label{predict:OLS-rt}
E_i^{\text{Retweets}} \sim  \text{OLS}(\text{Complexity}_i + \text{Valence}_i+ \text{Topic}_i+ \text{Author}_i)
\end{equation}
\begin{equation}
\label{predict:OLS-c}
E_i^{\text{Comments}} \sim  \text{OLS}(\text{Complexity}_i + \text{Valence}_i+ \text{Topic}_i+ \text{Author}_i)
\end{equation}

In our regression models, we include control variables for author characteristics, acknowledging that these attributes contribute to a tweet's overall popularity. However, our primary focus remains on the content features, as unpacking the impact of author attributes on unexpectedly high engagement, while interesting, lies beyond the scope of this study. The estimation results of our regression analyses are presented in Figure \ref{fig:regressionResult}. To ensure the robustness of our findings, we present two sets of estimates: one using all tweets with at least one retweet, like, and comment, and another using only tweets that have at least 10 engagements of one type. This dual approach allows us to verify that our results are not driven by outliers or low-engagement tweets. Notably, we find similar patterns in both versions, lending credence to the stability of our findings across different engagement thresholds.

\subsubsection{Summary of Key Findings:}

Our analysis reveals several key observations regarding the impact of content features on unexpected engagement patterns. In terms of topic, we find that news, politics, and business-related content tends to receive higher than expected retweets and comments. This finding aligns with Attention Economy Theory, suggesting that these topics may be perceived as particularly valuable or relevant information, prompting users to engage more deeply and share with their networks. In contrast, games and sports-related topics are more likely to receive unexpectedly high numbers of comments and likes, possibly indicating a higher degree of emotional investment or personal opinion-sharing among users interested in these subjects. Celebrity and movie-related topics usually garner higher than expected retweets and likes, while art topics primarily attract more likes than anticipated. These patterns suggest that different topics elicit varied forms of engagement, potentially serving as different types of signals in the social media ecosystem.

Regarding valence, our results indicate that tweets with a high degree of subjectivity tend to attract higher than expected likes. This could be interpreted through the lens of Signaling Theory as users employing likes to signal agreement or support for subjective opinions. Conversely, objective tweets primarily attract more retweets than predicted, possibly indicating that users are more inclined to share factual information with their networks. Positivity in a tweet's wording (represented as "sentiment" in our analysis) is associated with higher than expected likes and, for tweets with higher overall engagement, more comments than predicted. However, it's noteworthy that the effect sizes for subjectivity and positivity are substantially smaller than those for topics, suggesting that the subject matter of a tweet plays a more significant role in determining engagement patterns than its emotional tone.

In terms of textual complexity, we find that tweets including URLs and those with more complex wording or greater length tend to attract more retweets than expected. This could indicate that users are more likely to share content they perceive as informative or substantive. Tweets with URLs also receive more likes than expected, possibly reflecting user appreciation for content that provides additional context or information. Longer tweets are associated with increased comments, suggesting that more extensive content may provoke discussion or debate. Interestingly, concrete tweets (those using more tangible or specific language) primarily attract higher likes than expected, potentially indicating that users more readily engage with content that is clear and easily understood.

These findings offer valuable insights into the complex dynamics of social media engagement, illuminating how different content features can serve as signals and compete for user attention in unexpected ways. In the following section, we will discuss the implications of these results in greater depth, exploring how they contribute to our understanding of online social behavior and what they might mean for content creators, platform designers, and researchers studying social media dynamics.

\subsection{Model Validation and Extensions}
\add{
To ensure the robustness of our unexpectedness quotient and subsequent analyses, we employed several validation procedures. We conducted k-fold cross-validation by dividing our dataset into training and testing subsets (k=10), which allowed us to assess the stability of our findings across different partitions of data. This validation approach confirmed that our results are not artifacts of a particular data subset but represent consistent patterns across our dataset. For each fold, we computed the unexpectedness quotient using the same procedure described in Equations 1-6, ensuring methodological consistency while testing model stability. The cross-validation results demonstrated high consistency in both the direction and magnitude of effects across folds, with average coefficient variations below 5\% for our main predictors, supporting the reliability of our findings.

Beyond our primary linear model specifications, we also investigated potential non-linear relationships and interaction effects in our data. To capture non-linear effects, we introduced squared terms for each continuous variable (readability, sentiment, subjectivity, length, and concreteness) in our regression models. While these non-linear specifications captured additional variance in the data, the direction of our main effects remained qualitatively similar to those in the linear models. For clarity and interpretability, we present the linear model as our main result in Figure 6, though the non-linear components provide additional nuance to our understanding of how content features influence unexpected engagement patterns. The detailed model results for non-linear specifications are presented in the Appendix (Figure 7).

We further extended our models to examine interaction effects between content characteristics and topic domains, revealing several important insights. For example, tweet length contributes most significantly to increased unexpected likes for topics such as news and politics (with interaction coefficients of 0.3342 and 0.2764, respectively), but negatively impacts unexpected retweets and comments for these same topics. This suggests that longer political and news content may receive more acknowledgment without necessarily generating proportional sharing or discussion. Similar topic-specific effects emerged for other content features. Sentiment interacts meaningfully with topic domains—positive sentiment in news tweets significantly boosts unexpected likes compared to other topics (coefficient 0.1183), while substantially reducing unexpected retweets (coefficient -0.1076). These findings highlight the importance of considering domain context when interpreting content effects, as the same textual features may influence engagement differently across topic areas. The detailed model results for investigating interaction effects are also presented in the Appendix (Figures 8-10).

These validation procedures and model extensions strengthen confidence in our core findings while revealing additional complexities in how content characteristics drive unexpected engagement patterns. The persistence of similar directional effects across linear and non-linear specifications, alongside consistent cross-validation results, supports the validity of our unexpectedness quotient as a meaningful metric for analyzing asymmetric engagement patterns on social media. Meanwhile, the interaction effects between topics and content features provide valuable insights for both researchers and practitioners, suggesting that content optimization strategies should be tailored to specific domains rather than applied uniformly across different content types. When evaluating and designing systems to promote certain kinds of interaction, platform designers should consider these potential topic-specific trade-offs, as different textual features benefit different topics in distinct ways.}

\begin{table*}[t]

\caption{Content features associated with unexpected high comments, retweets, or likes.}
\begin{small}
\begin{tabular}{lllllll}
\toprule
                 & \textbf{Sentiment} & \textbf{Link} & \textbf{Subjectivity} & \textbf{Length} & \textbf{Concrete} & \textbf{Readability} \\
\midrule
\textbf{Unexpected High Comments} & Marginal              & No            &  Marginal             & Long            & Marginal          & Simple             \\
\textbf{Unexpected High Retweets} & Marginal              & Yes            & Objective            & Long            &  Marginal            & Complex               \\
\textbf{Unexpected High Likes}    & Marginal              & Yes            & Subjective             & Short            & Marginal         & Simple             
\\
\bottomrule
\end{tabular}
\end{small}
\label{tab:summary}
\end{table*}

\section{Discussion \& Conclusion}

Social media platforms offer users a variety of ways to interact with content, ranging from low-effort actions like liking or favoriting to more involved interactions such as retweeting and commenting. Through the lens of Signaling Theory, these different affordances can be viewed as signals of varying strength and cost \citep{donath2007signals, connelly2011signaling}. Likes represent low-cost signals of approval, while retweets and comments serve as stronger signals of endorsement and engagement, respectively. Simultaneously, the Attention Economy Theory frames these interactions as competing for users' limited attention resources \citep{davenport2001attention, webster2014marketplace}. Our study analyzes these three primary types of interactions on Twitter, utilizing an extensive set of content properties to quantify and investigate the determinants of unexpected engagement patterns. By examining these patterns through the dual perspectives of Signaling Theory and Attention Economy Theory, we offer insights that extend beyond academic interest to practical applications for social media practitioners and platform designers. Our focus on the impact of tweet content, which can be easily modified, provides actionable insights for achieving targeted engagement and diffusion within the constraints of the attention economy. Drawing from existing literature, we hypothesized that textual complexity, valence, and topic would influence a tweet's likelihood of receiving unexpectedly high engagement \citep{boyd2010tweet,meier2014more,levordashka2016s}, with each of these factors potentially serving as different types of signals or attention-capturing mechanisms.

\subsection{Summary of Key Findings}

Our empirical analysis of over half a million tweets has significantly advanced our understanding of unexpected interactions on Twitter and the content features that drive them. We found that textual complexity, topic, and valence play crucial roles in explaining unexpectedly high engagement of specific types on Twitter. These findings bridge the gap between general engagement studies and our more nuanced examination of unexpected interaction patterns, while also aligning with the principles of Signaling Theory and Attention Economy Theory \citep{goldhaber1997attention,mahmud2013will}. Our novel metric, the unexpectedness quotient, provides a robust tool for identifying content that breaks through the noise of social media, effectively capturing attention in a saturated information environment. This metric allows us to quantify how certain content features enable tweets to compete more effectively for user attention and elicit stronger engagement signals than would be expected based on overall popularity. Interestingly, our analysis revealed that unexpectedly high comments are more prevalent than higher-than-expected retweets or likes, suggesting that generating discussion may be a more attainable goal for content creators than achieving viral sharing. This aligns with Signaling Theory, as comments represent the strongest form of engagement signal, indicating a higher investment of time and cognitive resources by users \citep{oh2017trump}. From an Attention Economy perspective, this finding suggests that content which prompts discussion may be particularly valuable in capturing and retaining user attention, potentially leading to longer-term engagement and community building on the platform \citep{lampe2006face}.

The primacy of content topic in determining unexpected interactions underscores the importance of subject matter in capturing attention and eliciting specific types of signals from users. News and political content, for instance, tends to receive higher-than-predicted retweets and comments, likely due to its capacity to spark discussion and its perceived value in the attention economy. This finding echoes previous research on political engagement in online forums \citep{gonzalez2010structure, yardi2010dynamic} and can be interpreted as users signaling their engagement with current events or political ideologies through higher-cost interactions. In contrast, games and sports-related topics garnered unexpectedly high likes and comments, possibly due to the passionate nature of fan engagement, representing a different type of valued content in the attention economy \citep{zhang2016online}. These topic-based differences in engagement patterns highlight how different types of content compete for attention in distinct ways and elicit varied forms of signaling behavior from users. Art-related content primarily attracted unexpected likes, suggesting a tendency for users to show support for visual content without necessarily engaging in extensive discussion. This could be interpreted as users employing low-cost signals to indicate appreciation in a domain where detailed commentary might be perceived as requiring expertise \citep{weng2012competition}.

Our analysis of valence and textual complexity provides further insights into the interplay between content characteristics, user behavior, and the attention economy of social media. Subjective tweets promoted more likes than expected, while objective tweets gathered unexpectedly high comments. Through the lens of Signaling Theory, this pattern suggests that users employ different types of engagement as signals depending on the nature of the content \citep{sharma2012inferring}. Likes may serve as low-cost signals of agreement with subjective content, while comments on objective content represent a higher-cost signal of engagement, potentially indicating a desire to contribute to or challenge the presented information. These findings contribute to our understanding of how content characteristics influence user behavior and decision-making in the attention economy of social media \citep{stieglitz2013emotions,berger2012makes}. Regarding textual complexity, our results support the notion that informative content tends to attract more engagement, particularly in the form of retweets. This aligns with both Signaling Theory and Attention Economy Theory, as sharing substantive content may signal the user's access to valuable information, while also competing effectively for attention in a crowded information landscape \citep{tintarev2007effective, he2014predicting}.

\del{Our findings have significant implications for both platform designers and content producers navigating the attention economy of social media. Our results offer a practical guide for tailoring content to specific engagement goals, whether that's maximizing likes, retweets, or comments. Platform designers can use these insights to develop more nuanced engagement affordances that align with different content types and user intentions, potentially optimizing the distribution of attention across the platform \citep{gao2013modeling}. For content creators, understanding the relationship between content features and unexpected engagement can inform more effective strategies for breaking through the noise and capturing user attention. Moreover, our study underscores the importance of considering the interplay between content characteristics, user behavior, and platform dynamics in the evolving landscape of social media. As platforms continue to refine their algorithms and affordances, the ability to predict and generate unexpected engagement may become increasingly valuable in the attention economy \citep{anderson2007long}. Future research could extend this work by examining how these patterns evolve over time, investigating potential cultural differences in engagement behaviors, or exploring how unexpected engagement translates into broader measures of content impact or user influence within the framework of Signaling Theory and Attention Economy Theory.}

\add{
\subsection{Contributions to HCI and CSCW Research}

Our research advances the fields of Human-Computer Interaction and Computer-Supported Cooperative Work in several important ways. First, we contribute to understanding the complex human-computer interactions involved in content engagement by revealing how users employ different engagement mechanisms (likes, retweets, comments) as distinct signals with varying intentions and meanings. This insight helps platform designers create more nuanced interaction affordances that better align with users' communicative goals rather than treating all engagement as equivalent. By demonstrating that users make deliberate choices about engagement types based on content characteristics, our work challenges the prevailing design paradigm that treats engagement as a monolithic construct, suggesting instead that platforms should facilitate differentiated engagement pathways that acknowledge these distinct user intentions.

Second, our work extends CSCW's focus on computer-mediated collaboration by demonstrating how different content characteristics facilitate specific types of collaborative engagement behaviors. The finding that objective content systematically drives unexpected sharing while subjective content promotes acknowledgment behaviors reveals fundamental patterns in how content qualities shape collaborative information distribution. These patterns have important implications for designing collaborative information systems that aim to foster specific types of group interaction rather than maximizing overall engagement. For example, systems designed to promote collaborative information curation might prioritize features that enhance objectivity and verifiability, while platforms focused on community building might emphasize subjective expression and personalization.

Third, our ``unexpectedness quotient'' provides a novel analytical tool for CSCW researchers studying how platform design influences user behavior and content diffusion patterns. By focusing on the relationship between different engagement types rather than just their absolute values, this metric reveals patterns of community response that might otherwise be obscured in traditional engagement analyses. This approach aligns with current CSCW research priorities that emphasize understanding quality over quantity in online interactions. The unexpectedness quotient could be adapted for studying engagement patterns across diverse collaborative platforms, providing a methodological contribution that extends beyond our specific empirical findings.

Our findings also connect to ongoing HCI discourse around designing for meaningful engagement rather than engagement maximization. Recent work in HCI has criticized optimization approaches that prioritize engagement volume without considering its nature or quality~\cite{scissors2016,burke2016relationship}. Our results support this critical perspective by demonstrating that different content characteristics drive specific types of engagement that might be more or less valuable depending on context and purpose. This suggests potential redesigns of engagement metrics and recommendation systems that consider not just how much engagement content receives, but whether it receives the most appropriate types of engagement for its domain and purpose. By providing empirical evidence of systematic relationships between content features and engagement types, our work offers a foundation for designing systems that promote quality interactions aligned with both user intentions and platform goals, advancing the HCI community's evolving understanding of meaningful online engagement.

\subsection{Practical Implications for Platform Designers}

Our findings offer several concrete recommendations for platform designers seeking to create more engaging and meaningful social media experiences. Recommendation algorithms could be redesigned to consider unexpectedness patterns rather than merely optimizing for overall engagement volume. For instance, platforms could prioritize content that generates unexpected discussion (high comment-to-like ratios) in contexts where meaningful conversation adds value, such as news and political domains. This approach might involve weighting comments more heavily than likes or retweets when ranking news content, potentially surfacing thoughtful discussions rather than merely popular content.

Platform interfaces could also be modified to highlight different engagement metrics based on content domain - emphasizing comment counts for political content where discussion is valuable, while foregrounding sharing metrics for informational content where distribution provides public value. For example, Twitter's ``Top Tweets’’ algorithm could incorporate domain-specific engagement expectations, recognizing that a news article with an unexpectedly high comment rate might be more valuable to users than one with simply high overall engagement.

Our unexpectedness metrics also suggest opportunities for more sophisticated content evaluation frameworks that move beyond raw engagement counts. Platforms could develop multidimensional engagement quality scores that consider whether content is receiving the most appropriate types of engagement for its domain and purpose. Content receiving unexpected discussion might be flagged as potentially controversial but valuable, while content receiving unexpected sharing without discussion might suggest information of broad interest. These nuanced measures could help platforms identify and promote content that generates meaningful engagement rather than merely maximizing interaction volume, potentially improving user satisfaction and platform health.

\subsection{Strategic Guidelines for Content Creators}

Content creators can strategically leverage our findings to target specific engagement objectives rather than pursuing generic "engagement" goals. Those seeking to maximize information dissemination should focus on creating objective, informational content with moderate textual complexity and include relevant URLs to external resources, as these characteristics consistently drive unexpectedly high retweets across topics. Our results suggest that particularly for news, business, and political content, maintaining objectivity while providing sufficient informational depth maximizes sharing potential.

Conversely, creators aiming to build community through discussion should incorporate elements that prompt personal connection while maintaining sufficient complexity to warrant thoughtful responses. For political and news content, questions that invite perspective-sharing generate unexpectedly high comment rates, while for sports and games, content that references shared experiences and community identity triggers more discussion than predicted by other engagement metrics. This approach is particularly effective for community managers and brands seeking to foster active engagement rather than passive consumption.
For relationship-building and brand affinity objectives, subjective content with emotional resonance tends to generate unexpectedly high likes, particularly in domains like art, music, and entertainment. Creators seeking stronger parasocial connections should emphasize personal narrative and authentic emotional expression, especially positive sentiment in typically objective domains like business and technology where such emotion stands out as unexpected. This strategy aligns with our finding that subjective tweets consistently receive more likes than would be predicted based on their sharing or discussion rates.

These domain-specific strategies allow content creators to move beyond generic ``engagement optimization'' toward targeted approaches that align with specific communication objectives and audience development goals. Rather than pursuing all engagement types simultaneously, creators can strategically emphasize content characteristics that drive the specific engagement behaviors most valuable to their communication goals.}

\section{Limitations and Future Work}

While our study provides valuable insights into unexpected engagement patterns on Twitter, it is not without limitations, which open up several avenues for future research.

\addNew{
\subsection{Value of Correlational Findings}

Although the evidence provided by our study is correlational, it meets a high standard for methodological rigor and reliability. We subject every model to comprehensive robustness checks including random half-sample replications, quadratic and interaction terms, removal of high-leverage observations, and five-fold cross-validation. Across all validation procedures, the direction and relative magnitude of our main coefficients remain remarkably stable. We systematically test for omitted-variable bias by incorporating sequential blocks of controls for author reach, verification status, topical category, message length, sentiment, media flags, and external links. While these controls explain additional variance, they leave our focal content effects substantively unchanged. We minimize measurement error by analyzing engagement ratios rather than raw counts, which reduces noise from automated accounts and temporal artifacts. Our full-year sample ensures that short-term fluctuations from breaking news or seasonal events are averaged out, revealing persistent underlying relationships across hundreds of thousands of tweets and diverse content domains.

These stable patterns advance both practical and theoretical understanding. For practitioners, they provide reliable quantitative benchmarks when product decisions must be made before experimental evidence becomes available. A designer seeking to enhance information diffusion can confidently prioritize objective messages containing links, while reserving more speculative modifications for controlled testing. In competitive attention markets, delaying action until perfect causal identification is achieved imposes opportunity costs that may exceed the prediction error from well-validated correlational evidence. For theorists, our findings illuminate the distinct economic functions of different engagement types. The divergence between subjective language predicting unexpected likes and objective content predicting unexpected retweets reveals that users treat liking as a low-cost signal of personal agreement but view retweeting as a higher-commitment endorsement requiring substantive content. This refines Signaling Theory by demonstrating how signal costs emerge from the interaction between platform affordances and message characteristics, while extending Attention Economy frameworks by showing that users distribute limited cognitive resources across multiple engagement channels rather than along a single continuum. The consistency of these relationships across different model specifications, time periods, and engagement thresholds confirms they reflect genuine behavioral regularities rather than statistical artifacts, providing researchers and practitioners with actionable insights for developing targeted hypotheses and evidence-based strategies.
}

\subsection{External Factors and Contextual Influences}

\addNew{Engagement decisions are made in an environment where geopolitical shocks, cultural rituals, and platform rules all shape the marginal cost and benefit of each action. During 2018 the Twitter sphere absorbed the United States mid-term elections, the men’s World Cup, repeated North-Korea–U.S. summits, and early rumblings of the Brexit withdrawal deal. Each episode temporarily altered topical salience and may have raised the reputational return to commenting or resharing political content while crowding out attention to routine posts. Platform affordances evolved as well: the 280-character limit introduced in late-2017 settled into common use, the algorithmic ``Home'' timeline replaced strict reverse chronology for large segments of users, and a prompt encouraging quote-tweets appeared on many mobile clients. Such changes can shift the opportunity cost of writing versus retweeting and thereby reweight our unexpectedness measures even when content attributes stay constant. Although a full-year sample cushions against single-day spikes, it cannot fully absorb regime breaks that persist for weeks. Consequently, some residual variation in our models likely reflects these contextual forces rather than unobserved content traits.

Future work could integrate these external influences through event-study and structural-break designs that complement our cross-sectional approach. High-resolution time stamps make it feasible to estimate difference-in-differences models around exogenous shocks such as the API-policy tightening that followed the Cambridge Analytica hearings or the staged rollout of the algorithmic feed. Interacting content features with event-window indicators would reveal whether, for example, objective language becomes a stronger predictor of retweets when political stakes are salient. Researchers could add macro covariates\textemdash Google Trends indices, VIX volatility, or television news airtime\textemdash to capture shifts in collective attention that are orthogonal to individual tweets. On the design side, platform engineers could treat identified structural breaks as natural experiments: if the quote-tweet prompt raises the incremental retweet quotient for URL-rich messages, that affordance can be deployed selectively in knowledge-sharing contexts. Accounting explicitly for geopolitical events and platform interventions would therefore sharpen causal inference and deliver more precise ranking rules, while also stress-testing whether the content mechanisms documented here survive under markedly different information regimes.
}

\subsection{Cross-Platform Generalizability}
\addNew{The unexpectedness quotient travels well across digital ecosystems because it rests on two structural preconditions rather than on any Twitter‐specific quirk. First, a platform must offer distinct engagement actions that vary in visibility, effort, or reputational stake; second, those actions must be jointly observable at the post-level. When these conditions hold, the ``conditional expectation'' baseline that we estimate for comments, likes, and retweets can be re-estimated without alteration on Facebook, LinkedIn, or YouTube, where reactions, shares, and replies play analogous roles. Population mix and interface design will change coefficient magnitudes—Facebook skews older, LinkedIn more professional, YouTube comments persist longer—but the abnormal-ratio construction is scale-invariant. A marketer seeking Facebook posts that attract “too many” reshares relative to reactions can therefore fit platform-specific quantile regressions and compute the same quotient. By the same logic, LinkedIn engineers could surface articles whose unexpected repost score signals diffusion potential not captured by mere applause counts. Because the metric normalizes within-platform distributions, it remains valid across languages and regional markets so long as each engagement mode exhibits sufficient dispersion. In short, the framework requires only differentiated actions and granular data, both of which characterize most mature social networks.

Adaptation becomes more nuanced when engagement channels are fewer or qualitatively different. Reddit compresses positive feedback into a single up-vote, so the quotient would have to predict comment volume from time-series signals such as early vote velocity rather than from an alternative reaction type. Instagram and TikTok, where likes dominate and comments are sparse, invite a reformulation that asks whether a reel secures an unexpectedly high share-out or save rate conditional on view count and likes, with watch-time serving as an implicit baseline. Visual emphasis on these platforms also necessitates feature sets that capture color dynamics, motion complexity, and face detection in lieu of textual concreteness scores. Such adjustments preserve the core economic insight\textemdash that users incur incremental costs when moving from passive acknowledgment to public propagation\textemdash while honoring the engineering realities of each interface. A comparative study could embed the framework in a multilevel model with platform-specific slopes and a shared hyper-prior, testing whether the cost hierarchy of actions is universal or contingent on local norms. If slopes cluster tightly, designers gain license to adopt common ranking heuristics; if they diverge, platform-tailored algorithms become imperative. Either outcome deepens theory by specifying the boundary conditions under which signaling costs shape engagement, and the self-normalizing nature of our metric provides a ready foundation for that cross-platform synthesis.
}

\subsection{Temporal Limitations and Platform Evolution}
Our analysis uses data from 2018, which raises questions about the continued relevance of our findings given Twitter's evolution since then, including its rebranding as X and significant interface changes. We recognize that platform algorithms, affordances, and user behaviors evolve over time, potentially altering the absolute levels of engagement that different content receives. However, several factors suggest the underlying relationships we identify likely remain relevant despite these changes. Recent comparative studies by \citet{shahbaznezhad2021role} and \citet{moran2020message} demonstrate that while absolute engagement levels fluctuate over time, the relative relationships between content features and engagement types remain surprisingly consistent across platform iterations.

The content-engagement relationships we identify are tied to fundamental aspects of human communication and information processing rather than specific platform implementations. For instance, the tendency for objective content to receive more unexpected retweets likely reflects enduring human information-sharing behaviors rather than platform-specific features. Interface changes may alter how users engage, but content characteristics that drive unexpected engagement patterns (like informational value, emotional resonance, and topic domain) continue to influence user decisions about how to interact with content. Additionally, our focus on relationships between engagement types rather than absolute engagement levels makes our findings more resistant to temporal shifts in overall platform activity. While acknowledging these temporal limitations, we suggest that our findings reveal persistent patterns in how content characteristics drive engagement decisions, offering insights that remain valuable despite platform evolution.

\section{Ethics Considerations}

In conducting this study, we prioritized ethical considerations, with a particular focus on user privacy. Given the non-experimental nature of our research, we identified privacy protection as the most significant ethical concern and implemented comprehensive measures to minimize potential risks. We strictly adhered to the Twitter API Terms of Service\footnote{https://developer.twitter.com/en/developer-terms/agreement-and-policy}, ensuring that our data collection and usage methods were in full compliance with platform regulations. Our analysis was limited exclusively to tweets that were publicly available at the time of data collection, respecting users' privacy settings and intentions regarding content visibility.

To further safeguard user privacy, all example tweets included in this paper have been carefully paraphrased, preserving the essence of the content while significantly reducing the possibility of identifying individual authors. Our analysis and results are presented at an aggregate level, representing broader patterns of behavior among Twitter users who use hashtags, rather than focusing on specific individuals. This approach allows us to draw meaningful conclusions about user behavior without compromising individual privacy. By implementing these measures, we have strived to contribute valuable insights to the field of social media research while upholding high ethical standards and respecting the rights of individual users.

\begin{acks}
We would like to thank the anonymous reviewers and the Area Chair for their constructive comments that have helped improve this paper.
\end{acks}

\bibliographystyle{ACM-Reference-Format}
\bibliography{sample-base}


\begin{thebibliography}{67}


\ifx \showCODEN    \undefined \def \showCODEN     #1{\unskip}     \fi
\ifx \showISBNx    \undefined \def \showISBNx     #1{\unskip}     \fi
\ifx \showISBNxiii \undefined \def \showISBNxiii  #1{\unskip}     \fi
\ifx \showISSN     \undefined \def \showISSN      #1{\unskip}     \fi
\ifx \showLCCN     \undefined \def \showLCCN      #1{\unskip}     \fi
\ifx \shownote     \undefined \def \shownote      #1{#1}          \fi
\ifx \showarticletitle \undefined \def \showarticletitle #1{#1}   \fi
\ifx \showURL      \undefined \def \showURL       {\relax}        \fi
\providecommand\bibfield[2]{#2}
\providecommand\bibinfo[2]{#2}
\providecommand\natexlab[1]{#1}
\providecommand\showeprint[2][]{arXiv:#2}

\bibitem[Aldous et~al\mbox{.}(2019)]%
        {aldous2019view}
\bibfield{author}{\bibinfo{person}{Kholoud~Khalil Aldous},
  \bibinfo{person}{Jisun An}, {and} \bibinfo{person}{Bernard~J Jansen}.}
  \bibinfo{year}{2019}\natexlab{}.
\newblock \showarticletitle{View, like, comment, post: Analyzing user
  engagement by topic at 4 levels across 5 social media platforms for 53 news
  organizations}. In \bibinfo{booktitle}{\emph{Proceedings of the International
  AAAI Conference on Web and Social Media}}, Vol.~\bibinfo{volume}{13}.
  \bibinfo{pages}{47--57}.
\newblock


\bibitem[Angrist and Pischke(2008)]%
        {angrist2008mostly}
\bibfield{author}{\bibinfo{person}{Joshua~D Angrist} {and}
  \bibinfo{person}{J{\"o}rn-Steffen Pischke}.} \bibinfo{year}{2008}\natexlab{}.
\newblock \bibinfo{booktitle}{\emph{Mostly harmless econometrics: An
  empiricist's companion}}.
\newblock \bibinfo{publisher}{Princeton university press}.
\newblock


\bibitem[Bakshy et~al\mbox{.}(2011)]%
        {bakshy2011everyone}
\bibfield{author}{\bibinfo{person}{Eytan Bakshy}, \bibinfo{person}{Jake~M
  Hofman}, \bibinfo{person}{Winter~A Mason}, {and} \bibinfo{person}{Duncan~J
  Watts}.} \bibinfo{year}{2011}\natexlab{}.
\newblock \showarticletitle{Everyone's an influencer: quantifying influence on
  twitter}. In \bibinfo{booktitle}{\emph{Proceedings of the fourth ACM
  international conference on Web search and data mining}}.
  \bibinfo{pages}{65--74}.
\newblock


\bibitem[Berger and Milkman(2012)]%
        {berger2012makes}
\bibfield{author}{\bibinfo{person}{Jonah Berger} {and}
  \bibinfo{person}{Katherine~L Milkman}.} \bibinfo{year}{2012}\natexlab{}.
\newblock \showarticletitle{What makes online content viral?}
\newblock \bibinfo{journal}{\emph{Journal of marketing research}}
  \bibinfo{volume}{49}, \bibinfo{number}{2} (\bibinfo{year}{2012}),
  \bibinfo{pages}{192--205}.
\newblock


\bibitem[Bernstein et~al\mbox{.}(2013)]%
        {bernstein2013quantifying}
\bibfield{author}{\bibinfo{person}{Michael~S Bernstein}, \bibinfo{person}{Eytan
  Bakshy}, \bibinfo{person}{Moira Burke}, {and} \bibinfo{person}{Brian
  Karrer}.} \bibinfo{year}{2013}\natexlab{}.
\newblock \showarticletitle{Quantifying the invisible audience in social
  networks}. In \bibinfo{booktitle}{\emph{Proceedings of the SIGCHI conference
  on human factors in computing systems}}. \bibinfo{pages}{21--30}.
\newblock


\bibitem[boyd et~al\mbox{.}(2010)]%
        {boyd2010tweet}
\bibfield{author}{\bibinfo{person}{danah boyd}, \bibinfo{person}{Scott Golder},
  {and} \bibinfo{person}{Gilad Lotan}.} \bibinfo{year}{2010}\natexlab{}.
\newblock \showarticletitle{Tweet, tweet, retweet: Conversational aspects of
  retweeting on twitter}. In \bibinfo{booktitle}{\emph{2010 43rd Hawaii
  international conference on system sciences}}. IEEE, \bibinfo{pages}{1--10}.
\newblock


\bibitem[Brady et~al\mbox{.}(2017)]%
        {brady2017emotion}
\bibfield{author}{\bibinfo{person}{William~J Brady}, \bibinfo{person}{Julian~A
  Wills}, \bibinfo{person}{John~T Jost}, \bibinfo{person}{Joshua~A Tucker},
  {and} \bibinfo{person}{Jay~J Van~Bavel}.} \bibinfo{year}{2017}\natexlab{}.
\newblock \showarticletitle{Emotion shapes the diffusion of moralized content
  in social networks}.
\newblock \bibinfo{journal}{\emph{Proceedings of the National Academy of
  Sciences}} \bibinfo{volume}{114}, \bibinfo{number}{28}
  (\bibinfo{year}{2017}), \bibinfo{pages}{7313--7318}.
\newblock


\bibitem[Bruns(2021)]%
        {bruns2021after}
\bibfield{author}{\bibinfo{person}{Axel Bruns}.}
  \bibinfo{year}{2021}\natexlab{}.
\newblock \showarticletitle{After the ‘APIcalypse’: Social media platforms
  and their fight against critical scholarly research}.
\newblock \bibinfo{journal}{\emph{Disinformation and data lockdown on social
  platforms}} (\bibinfo{year}{2021}), \bibinfo{pages}{14--36}.
\newblock


\bibitem[Brysbaert et~al\mbox{.}(2014)]%
        {brysbaert2014concreteness}
\bibfield{author}{\bibinfo{person}{Marc Brysbaert}, \bibinfo{person}{Amy~Beth
  Warriner}, {and} \bibinfo{person}{Victor Kuperman}.}
  \bibinfo{year}{2014}\natexlab{}.
\newblock \showarticletitle{Concreteness ratings for 40 thousand generally
  known English word lemmas}.
\newblock \bibinfo{journal}{\emph{Behavior research methods}}
  \bibinfo{volume}{46}, \bibinfo{number}{3} (\bibinfo{year}{2014}),
  \bibinfo{pages}{904--911}.
\newblock


\bibitem[Burke and Kraut(2016)]%
        {burke2016relationship}
\bibfield{author}{\bibinfo{person}{Moira Burke} {and} \bibinfo{person}{Robert~E
  Kraut}.} \bibinfo{year}{2016}\natexlab{}.
\newblock \showarticletitle{The relationship between Facebook use and
  well-being depends on communication type and tie strength}.
\newblock \bibinfo{journal}{\emph{Journal of computer-mediated communication}}
  \bibinfo{volume}{21}, \bibinfo{number}{4} (\bibinfo{year}{2016}),
  \bibinfo{pages}{265--281}.
\newblock


\bibitem[Burke et~al\mbox{.}(2010)]%
        {burke2010social}
\bibfield{author}{\bibinfo{person}{Moira Burke}, \bibinfo{person}{Cameron
  Marlow}, {and} \bibinfo{person}{Thomas Lento}.}
  \bibinfo{year}{2010}\natexlab{}.
\newblock \showarticletitle{Social network activity and social well-being}. In
  \bibinfo{booktitle}{\emph{Proceedings of the SIGCHI conference on human
  factors in computing systems}}. \bibinfo{pages}{1909--1912}.
\newblock


\bibitem[Cha et~al\mbox{.}(2010)]%
        {cha2010measuring}
\bibfield{author}{\bibinfo{person}{Meeyoung Cha}, \bibinfo{person}{Hamed
  Haddadi}, \bibinfo{person}{Fabricio Benevenuto}, {and}
  \bibinfo{person}{Krishna Gummadi}.} \bibinfo{year}{2010}\natexlab{}.
\newblock \showarticletitle{Measuring user influence in twitter: The million
  follower fallacy}. In \bibinfo{booktitle}{\emph{Proceedings of the
  international AAAI conference on web and social media}},
  Vol.~\bibinfo{volume}{4}. \bibinfo{pages}{10--17}.
\newblock


\bibitem[Connelly et~al\mbox{.}(2011)]%
        {connelly2011signaling}
\bibfield{author}{\bibinfo{person}{Brian~L Connelly}, \bibinfo{person}{S~Trevis
  Certo}, \bibinfo{person}{R~Duane Ireland}, {and}
  \bibinfo{person}{Christopher~R Reutzel}.} \bibinfo{year}{2011}\natexlab{}.
\newblock \showarticletitle{Signaling theory: A review and assessment}.
\newblock \bibinfo{journal}{\emph{Journal of management}} \bibinfo{volume}{37},
  \bibinfo{number}{1} (\bibinfo{year}{2011}), \bibinfo{pages}{39--67}.
\newblock


\bibitem[Coscia(2017)]%
        {coscia2017popularity}
\bibfield{author}{\bibinfo{person}{Michele Coscia}.}
  \bibinfo{year}{2017}\natexlab{}.
\newblock \showarticletitle{Popularity spikes hurt future chances for viral
  propagation of protomemes}.
\newblock \bibinfo{journal}{\emph{Commun. ACM}} \bibinfo{volume}{61},
  \bibinfo{number}{1} (\bibinfo{year}{2017}), \bibinfo{pages}{70--77}.
\newblock


\bibitem[Davenport and Beck(2001)]%
        {davenport2001attention}
\bibfield{author}{\bibinfo{person}{Thomas~H Davenport} {and}
  \bibinfo{person}{John~C Beck}.} \bibinfo{year}{2001}\natexlab{}.
\newblock \showarticletitle{The attention economy}.
\newblock \bibinfo{journal}{\emph{Ubiquity}} \bibinfo{volume}{2001},
  \bibinfo{number}{May} (\bibinfo{year}{2001}), \bibinfo{pages}{1--es}.
\newblock


\bibitem[Donath(2007)]%
        {donath2007signals}
\bibfield{author}{\bibinfo{person}{Judith Donath}.}
  \bibinfo{year}{2007}\natexlab{}.
\newblock \showarticletitle{Signals in social supernets}.
\newblock \bibinfo{journal}{\emph{Journal of computer-mediated communication}}
  \bibinfo{volume}{13}, \bibinfo{number}{1} (\bibinfo{year}{2007}),
  \bibinfo{pages}{231--251}.
\newblock


\bibitem[Franck(2019)]%
        {franck2019economy}
\bibfield{author}{\bibinfo{person}{Georg Franck}.}
  \bibinfo{year}{2019}\natexlab{}.
\newblock \showarticletitle{The economy of attention}.
\newblock \bibinfo{journal}{\emph{Journal of sociology}} \bibinfo{volume}{55},
  \bibinfo{number}{1} (\bibinfo{year}{2019}), \bibinfo{pages}{8--19}.
\newblock


\bibitem[Freelon(2018)]%
        {freelon2018computational}
\bibfield{author}{\bibinfo{person}{Deen Freelon}.}
  \bibinfo{year}{2018}\natexlab{}.
\newblock \showarticletitle{Computational research in the post-API age}.
\newblock \bibinfo{journal}{\emph{Political Communication}}
  \bibinfo{volume}{35}, \bibinfo{number}{4} (\bibinfo{year}{2018}),
  \bibinfo{pages}{665--668}.
\newblock


\bibitem[Gilbert(2013)]%
        {gilbert2013widespread}
\bibfield{author}{\bibinfo{person}{Eric Gilbert}.}
  \bibinfo{year}{2013}\natexlab{}.
\newblock \showarticletitle{Widespread underprovision on reddit}. In
  \bibinfo{booktitle}{\emph{Proceedings of the 2013 conference on Computer
  supported cooperative work}}. \bibinfo{pages}{803--808}.
\newblock


\bibitem[Golder and Macy(2014)]%
        {golder2014digital}
\bibfield{author}{\bibinfo{person}{Scott~A Golder} {and}
  \bibinfo{person}{Michael~W Macy}.} \bibinfo{year}{2014}\natexlab{}.
\newblock \showarticletitle{Digital footprints: Opportunities and challenges
  for online social research}.
\newblock \bibinfo{journal}{\emph{Annual review of sociology}}
  \bibinfo{volume}{40}, \bibinfo{number}{1} (\bibinfo{year}{2014}),
  \bibinfo{pages}{129--152}.
\newblock


\bibitem[Goldhaber(1997)]%
        {goldhaber1997attention}
\bibfield{author}{\bibinfo{person}{Michael~H Goldhaber}.}
  \bibinfo{year}{1997}\natexlab{}.
\newblock \showarticletitle{The attention economy and the net}.
\newblock \bibinfo{journal}{\emph{First Monday}} (\bibinfo{year}{1997}).
\newblock


\bibitem[Gonzalez-Bailon et~al\mbox{.}(2010)]%
        {gonzalez2010structure}
\bibfield{author}{\bibinfo{person}{Sandra Gonzalez-Bailon},
  \bibinfo{person}{Andreas Kaltenbrunner}, {and} \bibinfo{person}{Rafael~E
  Banchs}.} \bibinfo{year}{2010}\natexlab{}.
\newblock \showarticletitle{The structure of political discussion networks: a
  model for the analysis of online deliberation}.
\newblock \bibinfo{journal}{\emph{Journal of Information Technology}}
  \bibinfo{volume}{25}, \bibinfo{number}{2} (\bibinfo{year}{2010}),
  \bibinfo{pages}{230--243}.
\newblock


\bibitem[Grevet et~al\mbox{.}(2014)]%
        {grevet2014managing}
\bibfield{author}{\bibinfo{person}{Catherine Grevet}, \bibinfo{person}{Loren~G
  Terveen}, {and} \bibinfo{person}{Eric Gilbert}.}
  \bibinfo{year}{2014}\natexlab{}.
\newblock \showarticletitle{Managing political differences in social media}. In
  \bibinfo{booktitle}{\emph{Proceedings of the 17th ACM conference on Computer
  supported cooperative work \& social computing}}.
  \bibinfo{pages}{1400--1408}.
\newblock


\bibitem[Hansen et~al\mbox{.}(2011)]%
        {hansen2011good}
\bibfield{author}{\bibinfo{person}{Lars~Kai Hansen}, \bibinfo{person}{Adam
  Arvidsson}, \bibinfo{person}{Finn~Aarup Nielsen}, \bibinfo{person}{Elanor
  Colleoni}, {and} \bibinfo{person}{Michael Etter}.}
  \bibinfo{year}{2011}\natexlab{}.
\newblock \showarticletitle{Good friends, bad news-affect and virality in
  twitter}. In \bibinfo{booktitle}{\emph{Future Information Technology: 6th
  International Conference, FutureTech 2011, Loutraki, Greece, June 28-30,
  2011, Proceedings, Part II}}. Springer, \bibinfo{pages}{34--43}.
\newblock


\bibitem[He et~al\mbox{.}(2014)]%
        {he2014predicting}
\bibfield{author}{\bibinfo{person}{Xiangnan He}, \bibinfo{person}{Ming Gao},
  \bibinfo{person}{Min-Yen Kan}, \bibinfo{person}{Yiqun Liu}, {and}
  \bibinfo{person}{Kazunari Sugiyama}.} \bibinfo{year}{2014}\natexlab{}.
\newblock \showarticletitle{Predicting the popularity of web 2.0 items based on
  user comments}. In \bibinfo{booktitle}{\emph{Proceedings of the 37th
  international ACM SIGIR conference on Research \& development in information
  retrieval}}. \bibinfo{pages}{233--242}.
\newblock


\bibitem[Huberman et~al\mbox{.}(2008)]%
        {huberman2008social}
\bibfield{author}{\bibinfo{person}{Bernardo~A Huberman},
  \bibinfo{person}{Daniel~M Romero}, {and} \bibinfo{person}{Fang Wu}.}
  \bibinfo{year}{2008}\natexlab{}.
\newblock \showarticletitle{Social networks that matter: Twitter under the
  microscope}.
\newblock \bibinfo{journal}{\emph{arXiv preprint arXiv:0812.1045}}
  (\bibinfo{year}{2008}).
\newblock


\bibitem[Hutto and Gilbert(2014)]%
        {hutto2014vader}
\bibfield{author}{\bibinfo{person}{Clayton~J Hutto} {and} \bibinfo{person}{Eric
  Gilbert}.} \bibinfo{year}{2014}\natexlab{}.
\newblock \showarticletitle{Vader: A parsimonious rule-based model for
  sentiment analysis of social media text}. In \bibinfo{booktitle}{\emph{Eighth
  international AAAI conference on weblogs and social media}}.
\newblock


\bibitem[Jungherr(2015)]%
        {jungherr2015analyzing}
\bibfield{author}{\bibinfo{person}{Andreas Jungherr}.}
  \bibinfo{year}{2015}\natexlab{}.
\newblock \showarticletitle{Analyzing political communication with digital
  trace data}.
\newblock \bibinfo{journal}{\emph{Cham, Switzerland: Springer}}
  (\bibinfo{year}{2015}).
\newblock


\bibitem[Kincaid et~al\mbox{.}(1975)]%
        {kincaid1975derivation}
\bibfield{author}{\bibinfo{person}{J~Peter Kincaid}, \bibinfo{person}{Robert~P
  Fishburne~Jr}, \bibinfo{person}{Richard~L Rogers}, {and}
  \bibinfo{person}{Brad~S Chissom}.} \bibinfo{year}{1975}\natexlab{}.
\newblock \bibinfo{booktitle}{\emph{Derivation of new readability formulas
  (automated readability index, fog count and flesch reading ease formula) for
  navy enlisted personnel}}.
\newblock \bibinfo{type}{{T}echnical {R}eport}. \bibinfo{institution}{Naval
  Technical Training Command Millington TN Research Branch}.
\newblock


\bibitem[Kwak et~al\mbox{.}(2010)]%
        {kwak2010twitter}
\bibfield{author}{\bibinfo{person}{Haewoon Kwak}, \bibinfo{person}{Changhyun
  Lee}, \bibinfo{person}{Hosung Park}, {and} \bibinfo{person}{Sue Moon}.}
  \bibinfo{year}{2010}\natexlab{}.
\newblock \showarticletitle{What is Twitter, a social network or a news
  media?}. In \bibinfo{booktitle}{\emph{Proceedings of the 19th international
  conference on World wide web}}. \bibinfo{pages}{591--600}.
\newblock


\bibitem[Lampe et~al\mbox{.}(2006)]%
        {lampe2006face}
\bibfield{author}{\bibinfo{person}{Cliff Lampe}, \bibinfo{person}{Nicole
  Ellison}, {and} \bibinfo{person}{Charles Steinfield}.}
  \bibinfo{year}{2006}\natexlab{}.
\newblock \showarticletitle{A Face (book) in the crowd: Social searching vs.
  social browsing}. In \bibinfo{booktitle}{\emph{Proceedings of the 2006 20th
  anniversary conference on Computer supported cooperative work}}.
  \bibinfo{pages}{167--170}.
\newblock


\bibitem[Lehmann et~al\mbox{.}(2012)]%
        {lehmann2012dynamical}
\bibfield{author}{\bibinfo{person}{Janette Lehmann}, \bibinfo{person}{Bruno
  Gon{\c{c}}alves}, \bibinfo{person}{Jos{\'e}~J Ramasco}, {and}
  \bibinfo{person}{Ciro Cattuto}.} \bibinfo{year}{2012}\natexlab{}.
\newblock \showarticletitle{Dynamical classes of collective attention in
  twitter}. In \bibinfo{booktitle}{\emph{Proceedings of the 21st international
  conference on World Wide Web}}. \bibinfo{pages}{251--260}.
\newblock


\bibitem[Levordashka et~al\mbox{.}(2016)]%
        {levordashka2016s}
\bibfield{author}{\bibinfo{person}{Ana Levordashka}, \bibinfo{person}{Sonja
  Utz}, {and} \bibinfo{person}{Renee Ambros}.} \bibinfo{year}{2016}\natexlab{}.
\newblock \showarticletitle{What’s in a like? Motivations for pressing the
  like button}. In \bibinfo{booktitle}{\emph{Proceedings of the International
  AAAI Conference on Web and Social Media}}, Vol.~\bibinfo{volume}{10}.
\newblock


\bibitem[Luo and Hancock(2020)]%
        {luo2020self}
\bibfield{author}{\bibinfo{person}{Mufan Luo} {and} \bibinfo{person}{Jeffrey~T
  Hancock}.} \bibinfo{year}{2020}\natexlab{}.
\newblock \showarticletitle{Self-disclosure and social media: motivations,
  mechanisms and psychological well-being}.
\newblock \bibinfo{journal}{\emph{Current opinion in psychology}}
  \bibinfo{volume}{31} (\bibinfo{year}{2020}), \bibinfo{pages}{110--115}.
\newblock


\bibitem[Macskassy and Michelson(2011)]%
        {macskassy2011people}
\bibfield{author}{\bibinfo{person}{Sofus Macskassy} {and}
  \bibinfo{person}{Matthew Michelson}.} \bibinfo{year}{2011}\natexlab{}.
\newblock \showarticletitle{Why do people retweet? anti-homophily wins the
  day!}. In \bibinfo{booktitle}{\emph{Proceedings of the International AAAI
  Conference on Web and Social Media}}, Vol.~\bibinfo{volume}{5}.
\newblock


\bibitem[Mahmud et~al\mbox{.}(2013)]%
        {mahmud2013will}
\bibfield{author}{\bibinfo{person}{Jalal Mahmud}, \bibinfo{person}{Jilin Chen},
  {and} \bibinfo{person}{Jeffrey Nichols}.} \bibinfo{year}{2013}\natexlab{}.
\newblock \showarticletitle{When will you answer this? estimating response time
  in twitter}. In \bibinfo{booktitle}{\emph{Proceedings of the International
  AAAI Conference on Web and Social Media}}, Vol.~\bibinfo{volume}{7}.
  \bibinfo{pages}{697--700}.
\newblock


\bibitem[Meier et~al\mbox{.}(2014)]%
        {meier2014more}
\bibfield{author}{\bibinfo{person}{Florian Meier}, \bibinfo{person}{David
  Elsweiler}, {and} \bibinfo{person}{Max Wilson}.}
  \bibinfo{year}{2014}\natexlab{}.
\newblock \showarticletitle{More than liking and bookmarking? towards
  understanding twitter favouriting behaviour}. In
  \bibinfo{booktitle}{\emph{Proceedings of the international AAAI conference on
  web and social media}}, Vol.~\bibinfo{volume}{8}.
\newblock


\bibitem[Moran et~al\mbox{.}(2020)]%
        {moran2020message}
\bibfield{author}{\bibinfo{person}{Gillian Moran}, \bibinfo{person}{Laurent
  Muzellec}, {and} \bibinfo{person}{Devon Johnson}.}
  \bibinfo{year}{2020}\natexlab{}.
\newblock \showarticletitle{Message content features and social media
  engagement: evidence from the media industry}.
\newblock \bibinfo{journal}{\emph{Journal of Product \& Brand Management}}
  \bibinfo{volume}{29}, \bibinfo{number}{5} (\bibinfo{year}{2020}),
  \bibinfo{pages}{533--545}.
\newblock


\bibitem[Naveed et~al\mbox{.}(2011)]%
        {naveed2011bad}
\bibfield{author}{\bibinfo{person}{Nasir Naveed}, \bibinfo{person}{Thomas
  Gottron}, \bibinfo{person}{J{\'e}r{\^o}me Kunegis}, {and}
  \bibinfo{person}{Arifah~Che Alhadi}.} \bibinfo{year}{2011}\natexlab{}.
\newblock \showarticletitle{Bad news travel fast: A content-based analysis of
  interestingness on twitter}. In \bibinfo{booktitle}{\emph{Proceedings of the
  3rd international web science conference}}. \bibinfo{pages}{1--7}.
\newblock


\bibitem[Oh and Kumar(2017)]%
        {oh2017trump}
\bibfield{author}{\bibinfo{person}{Chong Oh} {and} \bibinfo{person}{Savan
  Kumar}.} \bibinfo{year}{2017}\natexlab{}.
\newblock \showarticletitle{How trump won: the role of social media sentiment
  in political elections}.
\newblock  (\bibinfo{year}{2017}).
\newblock


\bibitem[Pancer and Poole(2016)]%
        {pancer2016popularity}
\bibfield{author}{\bibinfo{person}{Ethan Pancer} {and} \bibinfo{person}{Maxwell
  Poole}.} \bibinfo{year}{2016}\natexlab{}.
\newblock \showarticletitle{The popularity and virality of political social
  media: hashtags, mentions, and links predict likes and retweets of 2016 US
  presidential nominees’ tweets}.
\newblock \bibinfo{journal}{\emph{Social Influence}} \bibinfo{volume}{11},
  \bibinfo{number}{4} (\bibinfo{year}{2016}), \bibinfo{pages}{259--270}.
\newblock


\bibitem[Pang and Lee(2004)]%
        {pang2004sentimental}
\bibfield{author}{\bibinfo{person}{Bo Pang} {and} \bibinfo{person}{Lillian
  Lee}.} \bibinfo{year}{2004}\natexlab{}.
\newblock \showarticletitle{A sentimental education: Sentiment analysis using
  subjectivity summarization based on minimum cuts}.
\newblock \bibinfo{journal}{\emph{arXiv preprint cs/0409058}}
  (\bibinfo{year}{2004}).
\newblock


\bibitem[Paul et~al\mbox{.}(2019)]%
        {paul2019elites}
\bibfield{author}{\bibinfo{person}{Indraneil Paul}, \bibinfo{person}{Abhinav
  Khattar}, \bibinfo{person}{Ponnurangam Kumaraguru}, \bibinfo{person}{Manish
  Gupta}, {and} \bibinfo{person}{Shaan Chopra}.}
  \bibinfo{year}{2019}\natexlab{}.
\newblock \showarticletitle{Elites tweet? Characterizing the Twitter verified
  user network}. In \bibinfo{booktitle}{\emph{2019 IEEE 35th International
  Conference on Data Engineering Workshops (ICDEW)}}. IEEE,
  \bibinfo{pages}{278--285}.
\newblock


\bibitem[Posch et~al\mbox{.}(2013)]%
        {posch2013meaning}
\bibfield{author}{\bibinfo{person}{Lisa Posch}, \bibinfo{person}{Claudia
  Wagner}, \bibinfo{person}{Philipp Singer}, {and} \bibinfo{person}{Markus
  Strohmaier}.} \bibinfo{year}{2013}\natexlab{}.
\newblock \showarticletitle{Meaning as collective use: predicting semantic
  hashtag categories on twitter}. In \bibinfo{booktitle}{\emph{Proceedings of
  the 22nd International Conference on World Wide Web}}.
  \bibinfo{pages}{621--628}.
\newblock


\bibitem[Romero et~al\mbox{.}(2011)]%
        {romero2011differences}
\bibfield{author}{\bibinfo{person}{Daniel~M Romero}, \bibinfo{person}{Brendan
  Meeder}, {and} \bibinfo{person}{Jon Kleinberg}.}
  \bibinfo{year}{2011}\natexlab{}.
\newblock \showarticletitle{Differences in the mechanics of information
  diffusion across topics: idioms, political hashtags, and complex contagion on
  twitter}. In \bibinfo{booktitle}{\emph{Proceedings of the 20th international
  conference on World wide web}}. \bibinfo{pages}{695--704}.
\newblock


\bibitem[Scissors et~al\mbox{.}(2016)]%
        {scissors2016}
\bibfield{author}{\bibinfo{person}{Lauren Scissors}, \bibinfo{person}{Moira
  Burke}, {and} \bibinfo{person}{Steven Wengrovitz}.}
  \bibinfo{year}{2016}\natexlab{}.
\newblock \showarticletitle{What's in a Like? Attitudes and behaviors around
  receiving Likes on Facebook}. In \bibinfo{booktitle}{\emph{Proceedings of the
  19th acm conference on computer-supported cooperative work \& social
  computing}}. \bibinfo{pages}{1501--1510}.
\newblock


\bibitem[Sekimoto et~al\mbox{.}(2020)]%
        {sekimoto2020metrics}
\bibfield{author}{\bibinfo{person}{Kenshin Sekimoto},
  \bibinfo{person}{Yoshifumi Seki}, \bibinfo{person}{Mitsuo Yoshida}, {and}
  \bibinfo{person}{Kyoji Umemura}.} \bibinfo{year}{2020}\natexlab{}.
\newblock \showarticletitle{The metrics of keywords to understand the
  difference between retweet and like in each category}. In
  \bibinfo{booktitle}{\emph{2020 IEEE/WIC/ACM International Joint Conference on
  Web Intelligence and Intelligent Agent Technology (WI-IAT)}}. IEEE,
  \bibinfo{pages}{560--567}.
\newblock


\bibitem[Semaan et~al\mbox{.}(2015)]%
        {semaan2015navigating}
\bibfield{author}{\bibinfo{person}{Bryan Semaan}, \bibinfo{person}{Heather
  Faucett}, \bibinfo{person}{Scott Robertson}, \bibinfo{person}{Misa Maruyama},
  {and} \bibinfo{person}{Sara Douglas}.} \bibinfo{year}{2015}\natexlab{}.
\newblock \showarticletitle{Navigating imagined audiences: Motivations for
  participating in the online public sphere}. In
  \bibinfo{booktitle}{\emph{Proceedings of the 18th ACM Conference on Computer
  Supported Cooperative Work \& Social Computing}}.
  \bibinfo{pages}{1158--1169}.
\newblock


\bibitem[Semaan et~al\mbox{.}(2014)]%
        {semaan2014social}
\bibfield{author}{\bibinfo{person}{Bryan~C Semaan}, \bibinfo{person}{Scott~P
  Robertson}, \bibinfo{person}{Sara Douglas}, {and} \bibinfo{person}{Misa
  Maruyama}.} \bibinfo{year}{2014}\natexlab{}.
\newblock \showarticletitle{Social media supporting political deliberation
  across multiple public spheres: towards depolarization}. In
  \bibinfo{booktitle}{\emph{Proceedings of the 17th ACM conference on Computer
  supported cooperative work \& social computing}}.
  \bibinfo{pages}{1409--1421}.
\newblock


\bibitem[Shahbaznezhad et~al\mbox{.}(2021)]%
        {shahbaznezhad2021role}
\bibfield{author}{\bibinfo{person}{Hamidreza Shahbaznezhad},
  \bibinfo{person}{Rebecca Dolan}, {and} \bibinfo{person}{Mona Rashidirad}.}
  \bibinfo{year}{2021}\natexlab{}.
\newblock \showarticletitle{The role of social media content format and
  platform in users’ engagement behavior}.
\newblock \bibinfo{journal}{\emph{Journal of Interactive Marketing}}
  \bibinfo{volume}{53}, \bibinfo{number}{1} (\bibinfo{year}{2021}),
  \bibinfo{pages}{47--65}.
\newblock


\bibitem[Sharma et~al\mbox{.}(2012)]%
        {sharma2012inferring}
\bibfield{author}{\bibinfo{person}{Naveen~Kumar Sharma},
  \bibinfo{person}{Saptarshi Ghosh}, \bibinfo{person}{Fabricio Benevenuto},
  \bibinfo{person}{Niloy Ganguly}, {and} \bibinfo{person}{Krishna Gummadi}.}
  \bibinfo{year}{2012}\natexlab{}.
\newblock \showarticletitle{Inferring who-is-who in the Twitter social
  network}.
\newblock \bibinfo{journal}{\emph{ACM SIGCOMM Computer Communication Review}}
  \bibinfo{volume}{42}, \bibinfo{number}{4} (\bibinfo{year}{2012}),
  \bibinfo{pages}{533--538}.
\newblock


\bibitem[Simon(1996)]%
        {simon1996designing}
\bibfield{author}{\bibinfo{person}{Herbert~A Simon}.}
  \bibinfo{year}{1996}\natexlab{}.
\newblock \showarticletitle{Designing organizations for an information-rich
  world}.
\newblock \bibinfo{journal}{\emph{International Library of Critical Writings in
  Economics}}  \bibinfo{volume}{70} (\bibinfo{year}{1996}),
  \bibinfo{pages}{187--202}.
\newblock


\bibitem[Stieglitz and Dang-Xuan(2013)]%
        {stieglitz2013emotions}
\bibfield{author}{\bibinfo{person}{Stefan Stieglitz} {and}
  \bibinfo{person}{Linh Dang-Xuan}.} \bibinfo{year}{2013}\natexlab{}.
\newblock \showarticletitle{Emotions and information diffusion in social
  media—sentiment of microblogs and sharing behavior}.
\newblock \bibinfo{journal}{\emph{Journal of management information systems}}
  \bibinfo{volume}{29}, \bibinfo{number}{4} (\bibinfo{year}{2013}),
  \bibinfo{pages}{217--248}.
\newblock


\bibitem[Suh et~al\mbox{.}(2010)]%
        {suh2010want}
\bibfield{author}{\bibinfo{person}{Bongwon Suh}, \bibinfo{person}{Lichan Hong},
  \bibinfo{person}{Peter Pirolli}, {and} \bibinfo{person}{Ed~H Chi}.}
  \bibinfo{year}{2010}\natexlab{}.
\newblock \showarticletitle{Want to be retweeted? large scale analytics on
  factors impacting retweet in twitter network}. In
  \bibinfo{booktitle}{\emph{2010 IEEE Second International Conference on Social
  Computing}}. IEEE, \bibinfo{pages}{177--184}.
\newblock


\bibitem[Tan et~al\mbox{.}(2014)]%
        {tan2014effect}
\bibfield{author}{\bibinfo{person}{Chenhao Tan}, \bibinfo{person}{Lillian Lee},
  {and} \bibinfo{person}{Bo Pang}.} \bibinfo{year}{2014}\natexlab{}.
\newblock \showarticletitle{The effect of wording on message propagation:
  Topic-and author-controlled natural experiments on Twitter}.
\newblock \bibinfo{journal}{\emph{arXiv preprint arXiv:1405.1438}}
  (\bibinfo{year}{2014}).
\newblock


\bibitem[Tenenboim and Cohen(2015)]%
        {tenenboim2015prompts}
\bibfield{author}{\bibinfo{person}{Ori Tenenboim} {and}
  \bibinfo{person}{Akiba~A Cohen}.} \bibinfo{year}{2015}\natexlab{}.
\newblock \showarticletitle{What prompts users to click and comment: A
  longitudinal study of online news}.
\newblock \bibinfo{journal}{\emph{Journalism}} \bibinfo{volume}{16},
  \bibinfo{number}{2} (\bibinfo{year}{2015}), \bibinfo{pages}{198--217}.
\newblock


\bibitem[Tintarev and Masthoff(2007)]%
        {tintarev2007effective}
\bibfield{author}{\bibinfo{person}{Nava Tintarev} {and} \bibinfo{person}{Judith
  Masthoff}.} \bibinfo{year}{2007}\natexlab{}.
\newblock \showarticletitle{Effective explanations of recommendations:
  user-centered design}. In \bibinfo{booktitle}{\emph{Proceedings of the 2007
  ACM conference on Recommender systems}}. \bibinfo{pages}{153--156}.
\newblock


\bibitem[Wang et~al\mbox{.}(2014)]%
        {wang2014whispers}
\bibfield{author}{\bibinfo{person}{Gang Wang}, \bibinfo{person}{Bolun Wang},
  \bibinfo{person}{Tianyi Wang}, \bibinfo{person}{Ana Nika},
  \bibinfo{person}{Haitao Zheng}, {and} \bibinfo{person}{Ben~Y Zhao}.}
  \bibinfo{year}{2014}\natexlab{}.
\newblock \showarticletitle{Whispers in the dark: analysis of an anonymous
  social network}. In \bibinfo{booktitle}{\emph{Proceedings of the 2014
  conference on internet measurement conference}}. \bibinfo{pages}{137--150}.
\newblock


\bibitem[Wang et~al\mbox{.}(2016)]%
        {wang2016using}
\bibfield{author}{\bibinfo{person}{Yuan Wang}, \bibinfo{person}{Jie Liu},
  \bibinfo{person}{Yalou Huang}, {and} \bibinfo{person}{Xia Feng}.}
  \bibinfo{year}{2016}\natexlab{}.
\newblock \showarticletitle{Using hashtag graph-based topic model to connect
  semantically-related words without co-occurrence in microblogs}.
\newblock \bibinfo{journal}{\emph{IEEE Transactions on Knowledge and Data
  Engineering}} \bibinfo{volume}{28}, \bibinfo{number}{7}
  (\bibinfo{year}{2016}), \bibinfo{pages}{1919--1933}.
\newblock


\bibitem[Webster(2014)]%
        {webster2014marketplace}
\bibfield{author}{\bibinfo{person}{James~G Webster}.}
  \bibinfo{year}{2014}\natexlab{}.
\newblock \bibinfo{booktitle}{\emph{The marketplace of attention: How audiences
  take shape in a digital age}}.
\newblock \bibinfo{publisher}{Mit Press}.
\newblock


\bibitem[Weng et~al\mbox{.}(2012)]%
        {weng2012competition}
\bibfield{author}{\bibinfo{person}{Lilian Weng}, \bibinfo{person}{Alessandro
  Flammini}, \bibinfo{person}{Alessandro Vespignani}, {and}
  \bibinfo{person}{Fillipo Menczer}.} \bibinfo{year}{2012}\natexlab{}.
\newblock \showarticletitle{Competition among memes in a world with limited
  attention}.
\newblock \bibinfo{journal}{\emph{Scientific reports}} \bibinfo{volume}{2},
  \bibinfo{number}{1} (\bibinfo{year}{2012}), \bibinfo{pages}{335}.
\newblock


\bibitem[Yang et~al\mbox{.}(2012)]%
        {yang2012we}
\bibfield{author}{\bibinfo{person}{Lei Yang}, \bibinfo{person}{Tao Sun},
  \bibinfo{person}{Ming Zhang}, {and} \bibinfo{person}{Qiaozhu Mei}.}
  \bibinfo{year}{2012}\natexlab{}.
\newblock \showarticletitle{We know what@ you\# tag: does the dual role affect
  hashtag adoption?}. In \bibinfo{booktitle}{\emph{Proceedings of the 21st
  international conference on World Wide Web}}. \bibinfo{pages}{261--270}.
\newblock


\bibitem[Yardi and Boyd(2010)]%
        {yardi2010dynamic}
\bibfield{author}{\bibinfo{person}{Sarita Yardi} {and} \bibinfo{person}{Danah
  Boyd}.} \bibinfo{year}{2010}\natexlab{}.
\newblock \showarticletitle{Dynamic debates: An analysis of group polarization
  over time on twitter}.
\newblock \bibinfo{journal}{\emph{Bulletin of science, technology \& society}}
  \bibinfo{volume}{30}, \bibinfo{number}{5} (\bibinfo{year}{2010}),
  \bibinfo{pages}{316--327}.
\newblock


\bibitem[Yu et~al\mbox{.}(2024)]%
        {yujiangdhillon2024}
\bibfield{author}{\bibinfo{person}{Yulin Yu}, \bibinfo{person}{Julie Jiang},
  {and} \bibinfo{person}{Paramveer~S. Dhillon}.}
  \bibinfo{year}{2024}\natexlab{}.
\newblock \showarticletitle{Characterizing the Structure of Online
  Conversations Across Reddit}.
\newblock \bibinfo{journal}{\emph{Proc. ACM Hum.-Comput. Interact.}}
  \bibinfo{volume}{8}, \bibinfo{number}{CSCW2}, Article
  \bibinfo{articleno}{374} (\bibinfo{date}{Nov.} \bibinfo{year}{2024}),
  \bibinfo{numpages}{23}~pages.
\newblock
\href{https://doi.org/10.1145/3686913}{doi:\nolinkurl{10.1145/3686913}}


\bibitem[Zhang and Counts(2015)]%
        {zhang2015modeling}
\bibfield{author}{\bibinfo{person}{Amy~X Zhang} {and} \bibinfo{person}{Scott
  Counts}.} \bibinfo{year}{2015}\natexlab{}.
\newblock \showarticletitle{Modeling ideology and predicting policy change with
  social media: Case of same-sex marriage}. In
  \bibinfo{booktitle}{\emph{proceedings of the 33rd Annual ACM conference on
  human factors in computing systems}}. \bibinfo{pages}{2603--2612}.
\newblock


\bibitem[Zhang et~al\mbox{.}(2018)]%
        {zhang2018characterizing}
\bibfield{author}{\bibinfo{person}{Justine Zhang}, \bibinfo{person}{Cristian
  Danescu-Niculescu-Mizil}, \bibinfo{person}{Christina Sauper}, {and}
  \bibinfo{person}{Sean~J Taylor}.} \bibinfo{year}{2018}\natexlab{}.
\newblock \showarticletitle{Characterizing online public discussions through
  patterns of participant interactions}.
\newblock \bibinfo{journal}{\emph{Proceedings of the ACM on Human-Computer
  Interaction}} \bibinfo{volume}{2}, \bibinfo{number}{CSCW}
  (\bibinfo{year}{2018}), \bibinfo{pages}{1--27}.
\newblock


\bibitem[Zhang and Mao(2016)]%
        {zhang2016online}
\bibfield{author}{\bibinfo{person}{Jing Zhang} {and} \bibinfo{person}{En Mao}.}
  \bibinfo{year}{2016}\natexlab{}.
\newblock \showarticletitle{From online motivations to ad clicks and to
  behavioral intentions: An empirical study of consumer response to social
  media advertising}.
\newblock \bibinfo{journal}{\emph{Psychology \& Marketing}}
  \bibinfo{volume}{33}, \bibinfo{number}{3} (\bibinfo{year}{2016}),
  \bibinfo{pages}{155--164}.
\newblock


\end{thebibliography}
\clearpage

\appendix

\section{Tweet Examples (paraphrased)} \label{appendix}
\begin{table*}[h]
\centering
\scriptsize               
\resizebox{\textwidth}{!}{%
\begin{tabular}{p{2cm}p{8cm}r r r}
\toprule
\textbf{Type} & \textbf{Example} & \textbf{Like} & \textbf{RT} & \textbf{Cmt} \\ \midrule

\multirow{5}{*}{\parbox{2cm}{\raggedright
Politics/news tweets with unexpectedly high comments}}
 & (1) Thanks to Trump, there are way more jobs than work and fewer people on food stamps, and the unemployment rates of African Americans, Hispanic and females are historically low. It is pretty obvious that he cares about the poor more than Obama. & 3115 & 1390 & 6661 \\
 & (2) Breaking: Trump asks supporters to vote on a logo for “Space Force” gear. & 216 & 141 & 627 \\
 & (3) Sean Hannity of Fox News claimed that the “Deep State” put a “hit” out on Trump using the anonymous New York Times op-ed. & 43 & 26 & 145 \\
 & (4) Maxine Waters, Trump’s “Messenger Boy”, slams Nunes for closing Russia probe. & 282 & 85 & 384 \\ \addlinespace

\multirow{5}{*}{\parbox{2cm}{\raggedright
Politics/news tweets with unexpectedly high retweets}}
 & (1) It is so sad that radical anti-USA, Israel, Russia … Islamist forces killed our people for supporting Trump. & 12 & 188 & 1 \\
 & (2) Is Brexit good for business? [A poll] & 83 & 355 & 4 \\
 & (3) Explosive analysis thread shows that Israel and five other nations supported Trump to win illegally in the 2016 election. & 7 & 88 & 1 \\
 & (4) Breaking: There was a shooting at a Maryland high school; the school was on lockdown. & 17 & 75 & 2 \\ \addlinespace

\multirow{5}{*}{\parbox{2cm}{\raggedright
Games/sports tweets with unexpectedly high comments}}
 & (1) Epic Games reported that Fortnite made \$3 billion in 2018. How much did you spend on the game? & 221 & 20 & 102 \\
 & (2) Will you be playing God of War in four days? [A poll] & 35 & 19 & 69 \\
 & (3) Who are you backing to win the 2018 World Cup? Comment below with the corresponding emoji. & 28 & 1 & 122 \\
 & (4) Germany has been knocked out of the World Cup. Sum up your reaction with a GIF below. & 229 & 21 & 555 \\ \addlinespace

\multirow{5}{*}{\parbox{2cm}{\raggedright
Games/sports tweets with unexpectedly high likes}}
 & (1) Ross coloured Kassandra in Assassin’s Creed Photo Mode. & 427 & 19 & 1 \\
 & (2) Mercedes-Benz is ready to thrill fans at the 2018 F1 (Apr 8, 8:40 PM IST). & 868 & 48 & 2 \\
 & (3) Well done! You are incredible! [A photo of Tottenham Hotspur at UCL] & 3168 & 132 & 9 \\
 & (4) Dana White looks chuffed at WWE WrestleMania 34. & 239 & 7 & 2 \\ \addlinespace

\multirow{5}{*}{\parbox{2cm}{\raggedright
Art-related tweets with unexpectedly high likes}}
 & (1) It is raining. [Four photographs of the rain] & 650 & 25 & 3 \\
 & (2) Good morning—busy day ahead. Here is page 5 of my Inktober artwork. & 217 & 3 & 1 \\
 & (3) My hairstyle yesterday. [Instagram link] & 1491 & 68 & 1 \\
 & (4) My Inktober artwork: [Title]. [A drawing] & 264 & 8 & 2 \\
 & (5) Happy Birthday, YMS Adum! [A fan-art drawing] & 304 & 12 & 2 \\

\bottomrule
\end{tabular}}%
\caption{Tweets (paraphrased) illustrating key findings.}
\label{tab:discussion_examples_1}
\end{table*}

\begin{table*}[h]
\centering
\scriptsize                
\resizebox{\textwidth}{!}{%
\begin{tabular}{p{2cm}p{8cm}r r r}
\toprule
\textbf{Type} & \textbf{Example} & \textbf{Like} & \textbf{RT} & \textbf{Cmt} \\ \midrule

\multirow{5}{*}{\parbox{2cm}{\raggedright
Subjective tweets with unexpectedly high likes}}
 & (1) We are so proud of Shruti Haasan, who has just reached 8 million followers on Instagram. Congratulations to her! [A poster of Shruti Haasan] & 1086 &  39 &  3 \\
 & (2) Conor Dwyer is my favorite person in this world! [A link to an Instagram post] &  386 &   9 &  1 \\
 & (3) Tonight, I feel that I am so pretty. [Four selfies] & 10026 & 258 & 80 \\
 & (4) He is my dearest Christmas pig. I'll use this as my new profile picture. [A picture of Shane Dawson, a YouTuber] &  703 &  21 &  2 \\ \addlinespace

\multirow{5}{*}{\parbox{2cm}{\raggedright
Long, hard-to-read tweets with unexpectedly high retweets}}
 & (1) The countries with the most tweets having the \#MTVBRKPOPBTS hashtag in the past hour are (ranked by number of tweets): 1. South Korea, 2. Indonesia, … 15. Spain. & 102 & 308 &  5 \\
 & (2) Indigenous People of Biafra: Biafra Education Awareness Day (B.E.A.D) lectures will be held at Goldsmiths, University of London … &  10 &  53 &  1 \\

\bottomrule
\end{tabular}}%
\caption{Tweets (paraphrased) illustrating key findings.}
\label{tab:discussion_examples_2}
\end{table*}

\begin{figure*}[t]
\centering
\includegraphics[width=0.9\textwidth]{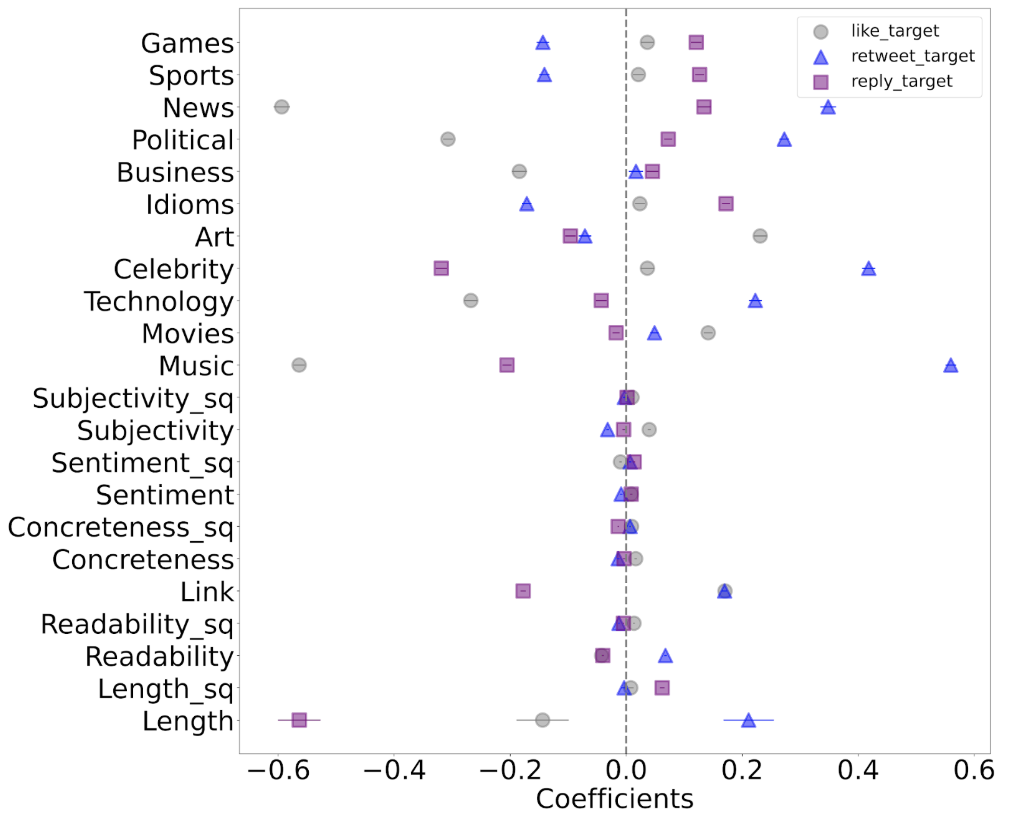}
\caption{OLS Regression estimation results for non-linear effect. The panel shows point estimates for equation  (\ref{predict:OLS-like}) (\ref{predict:OLS-rt}) and (\ref{predict:OLS-c}) using dataset with tweets with at least one like, retweet, and comment. {\it Note:} 1) Heteroskedasticity Robust standard errors are shown in parenthesis next to the coefficient.  2) The coefficients of the author features are omitted. 3) Error bar indicates 95\% confidential interval. 4) $_\text{sq}$ indicates the squared term of an independent variable. For example, $\text{length}_{sq}$ represents the coefficient of the squared term of the variable length in the regression model.}
\label{fig:regressionResult1}
\end{figure*}

\begin{figure*}[t]
\centering
\includegraphics[width=0.7\textwidth]{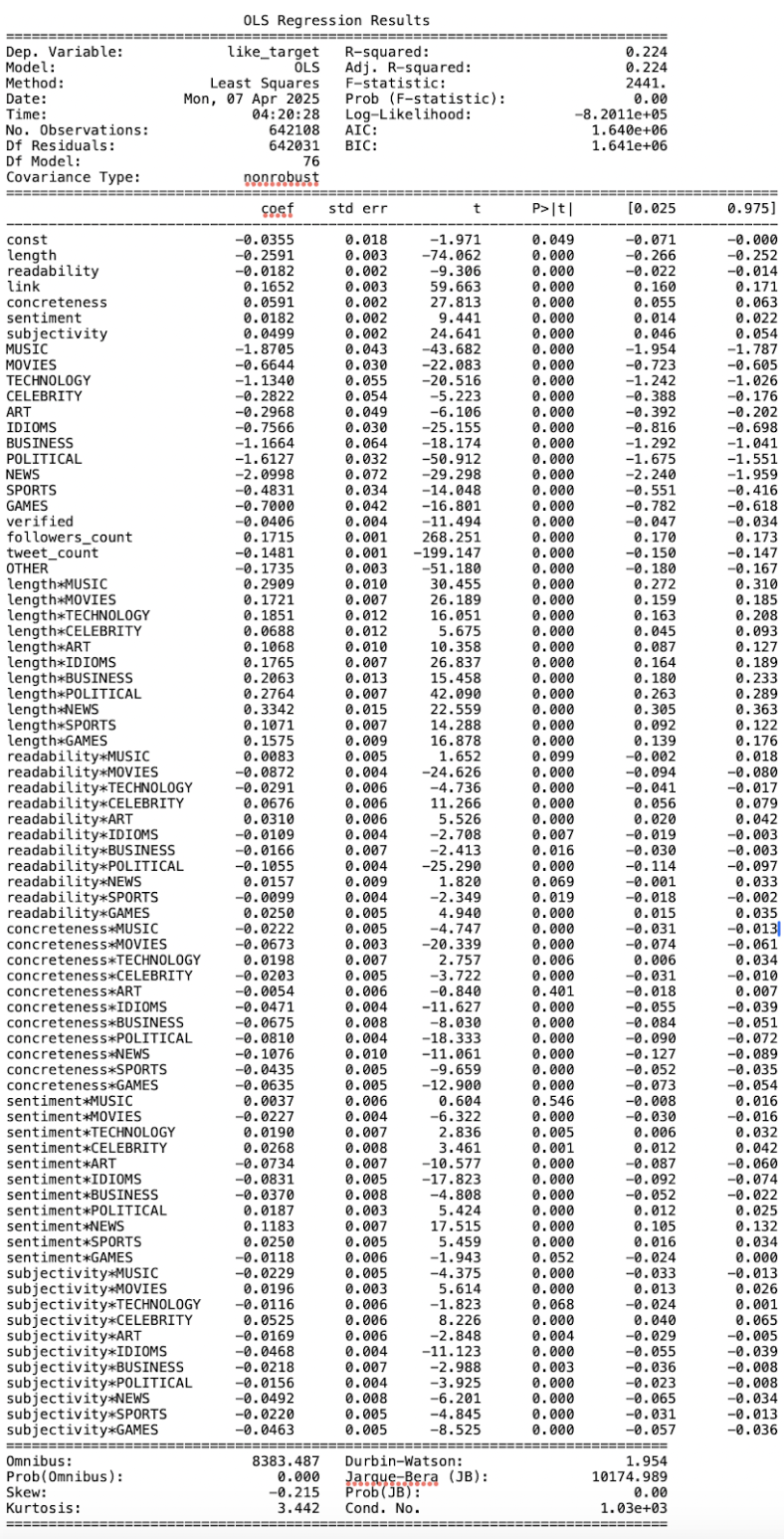}
\caption{OLS regression results examining the interaction effect on the outcome variable: unexpected likes.}
\label{fig:regressionResult1}
\end{figure*}

\begin{figure*}[t]
\centering
\includegraphics[width=0.7\textwidth]{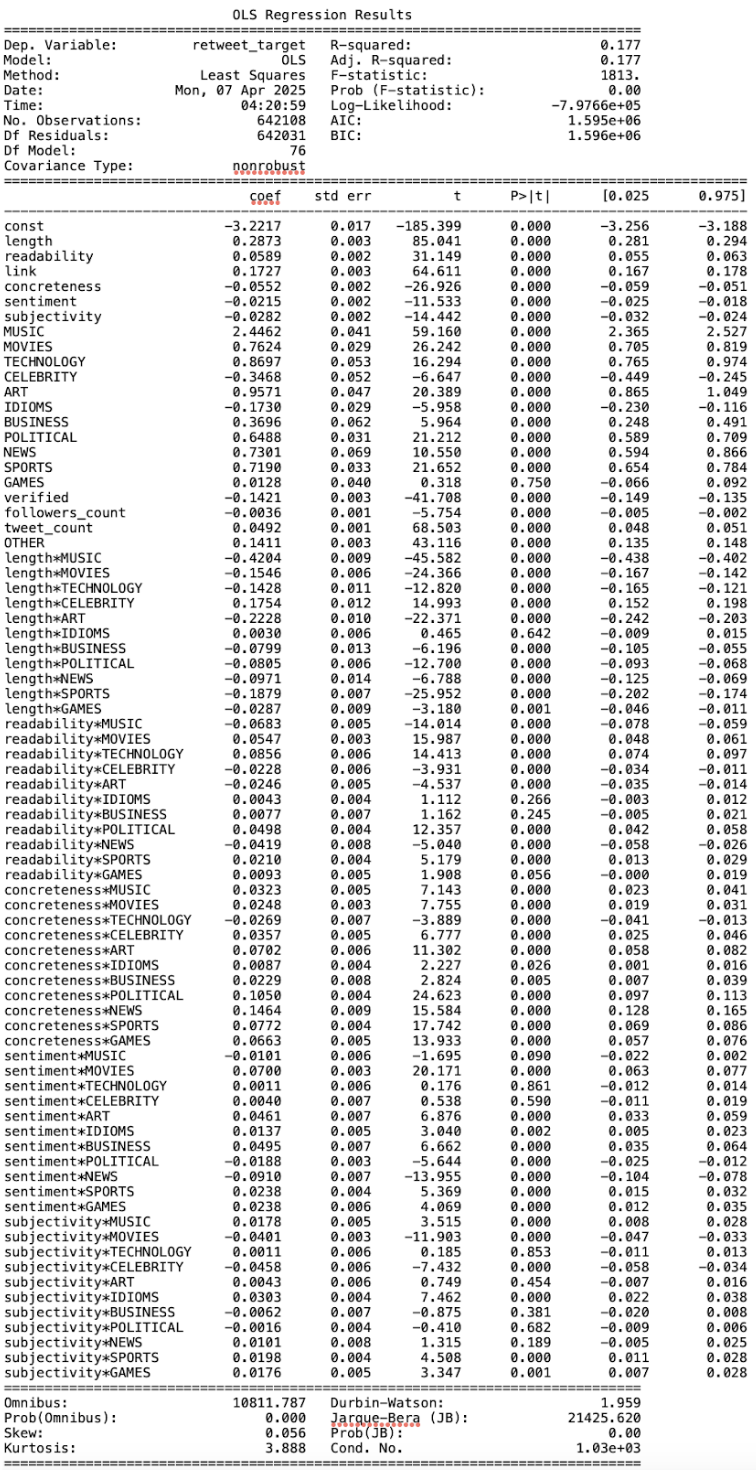}
\caption{OLS regression results examining the interaction effect on the outcome variable: unexpected retweet.}
\label{fig:regressionResult1}
\end{figure*}

\begin{figure*}[t]
\centering
\includegraphics[width=0.7\textwidth]{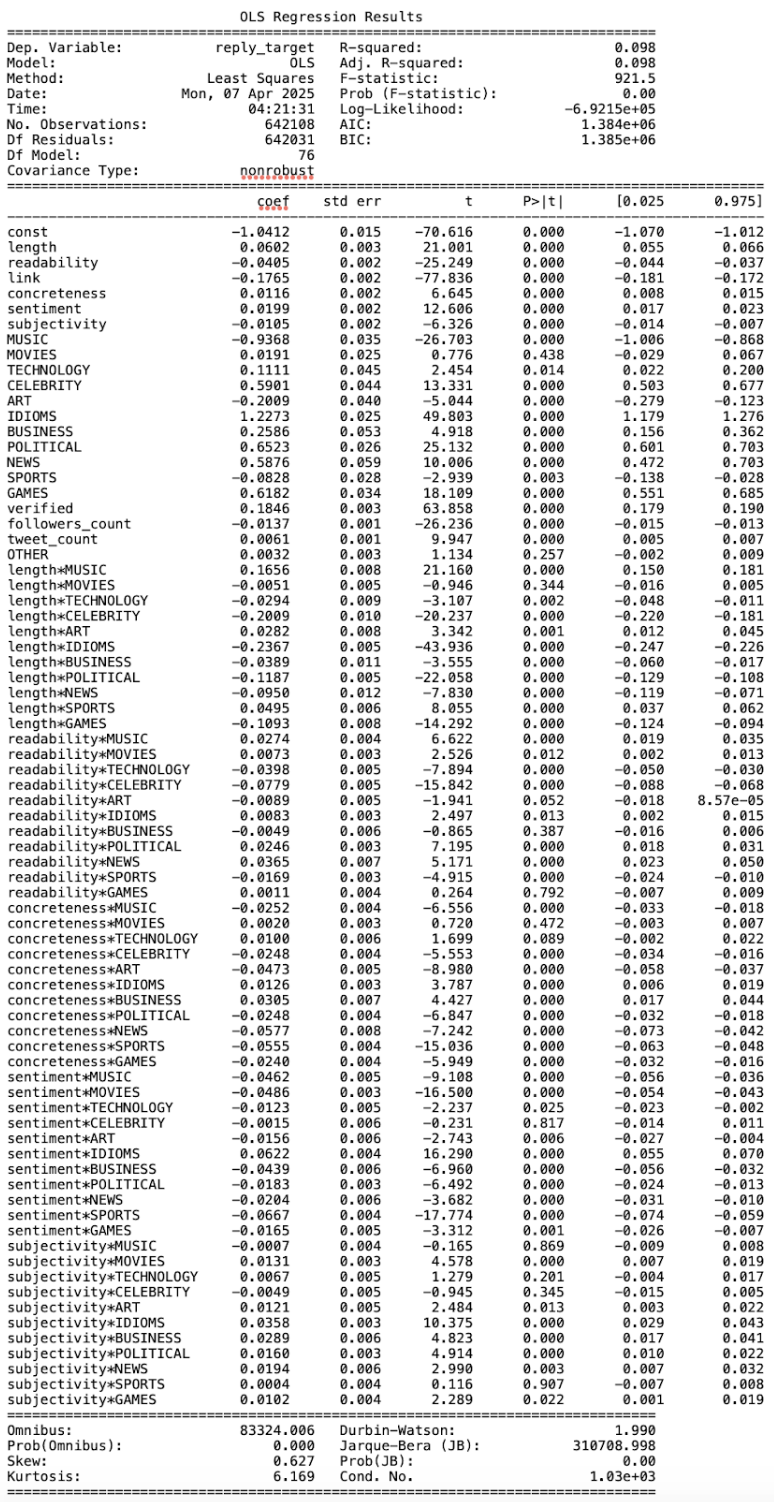}
\caption{OLS regression results examining the interaction effect on the outcome variable: unexpected comment.}
\label{fig:regressionResult1}
\end{figure*}

\end{document}